\documentclass[12pt]{JHEP} 
\usepackage{epsfig}
\newcommand\fverb{\setbox\pippobox=\hbox\bgroup\verb}
\newcommand\fverbdo{\egroup\medskip\noindent%
			\fbox{\unhbox\pippobox}\ }
\newcommand\fverbit{\egroup\item[\fbox{\unhbox\pippobox}]}
\newbox\pippobox

\def\be{\begin{equation}}
\def\ee{\end{equation}}
\def\ba{\begin{array}{l}}
\def\ea{\end{array}}
\def\bea{\begin{eqnarray}}
\def\eea{\end{eqnarray}}
\def\eq#1{(\ref{#1})}

\def\del{\partial}

\def\gap#1{\vspace{#1 ex}}

\def\half{\frac{1}{2}}

\def\x{{\bar x}}
\def\n{\mathtt{n}}

\def\I{{\mathbf I}}
\def\K{{\mathbf K}}

\def\V{\mathtt{V}}

\def\myitem{\gap3\noindent}

\def\nn{\nonumber}

\def\ket#1{| #1 \rangle}
\title{String bits in small radius AdS and weakly\\ coupled
${\cal N}=4$ Super Yang-Mills Theory: I}
\author{
Avinash Dhar, Gautam Mandal  and Spenta R. Wadia\\
{\it Department of Theoretical Physics,\\
Tata Institute of Fundamental Research,\\ 
Homi Bhabha Road, Mumbai 400 005, India.}
\\
E-mail: \email{adhar@theory.tifr.res.in,
mandal@theory.tifr.res.in, wadia@theory.tifr.res.in}}

\preprint{\hepth{0304062}\\
TIFR/TH/03-09}

\abstract{
We study light-cone gauge quantization of IIB strings in
AdS$_5 \times$ S$^5$ for small radius in Poincare coordinates. A
picture of strings made up of noninteracting bits emerges in the zero
radius limit. In this limit, each bit behaves like a superparticle
moving in the AdS$_5 \times$ S$^5$ background, carrying appropriate
representations of the super conformal group PSU$(2,2|4)$. The
standard Hamiltonian operator which causes evolution in the light-cone
time has continuous eigenvalues and provides a basis of states which
is not suitable for comparing with the dual super Yang-Mills
theory. However, there exist operators in the light-cone gauge which
have discrete spectra and can be used to label the states.  We obtain
the spectrum of single bit states and construct multi-bit states in this
basis. There are difficulties in the construction of string states
from the multi-bit states, which we discuss.  A non-zero value of the
radius introduces interactions between the bits and the spectrum of
multi-bit states gets modified. We compute the leading perturbative
corrections at small radius for a few simple cases.  Potential
divergences in the perturbative corrections, arising from strings near
the boundary, cancel.  This encourages us to believe that our
perturbative treatment could provide a framework for a rigorous and
detailed testing of the AdS/CFT conjecture, once the difficulties in
the construction of string states are resolved.  } \keywords{String
theory, Gauge theory} \dedicated{Dedicated to the memory of Bunji
Sakita.}

\begin{document}
\section{Introduction} 

The connection between $4$-dim. gauge theories and string theory is an
idea with a long history that goes back to attempts to discover
strings and string dynamics in gauge theories. A part of this history,
in connection with the 1/N expansion, can be found in
\cite{Brezin:eb}. See also \cite{Polyakov:ca,Polyakov:1998ju}.
 The hope underlying such attempts was
to guarantee consistent string dynamics in the context of a theory in
4 spacetime dimensions rather than in 10 or 26 spacetime
dimensions. The main motivation for finding a description of gauge
theories in terms of a string theory was to solve the problem of quark
confinement. This problem remains unsolved even today.
For a semi-popular
account of this problem see \cite{Current}.

It took more than two decades to arrive at a precise formulation of
this idea in the form of the AdS/CFT correspondence proposed by
Maldacena \cite{Mal97}. The most studied case of this correspondence is the
duality between $N=4$ supersymmetric Yang-Mills theory in 4-dimensions
and type IIB string theory on AdS$_5 \times$ S$^5$.  The clue came
from the seemingly unrelated problem of understanding thermal
properties of a class of black holes in terms of D-branes\footnote{see
\cite{DMW} for a recent review of this subject.}.

The AdS/CFT correspondence has been precisely formulated in the limit
when the 't Hooft coupling $\lambda=g_{YM}^2 N$ is large
\cite{GKP,W}. On the AdS side this means that $R^2=\alpha'\sqrt\lambda$
is large and the dual type IIB string theory is approximated by
supergravity. A correspondence between local gauge-invariant operators
of the gauge theory and local supergravity fields has been extensively
discussed in the literature \cite{MAGOO}. Besides the local operators,
the correspondence has also been studied for gauge-invariant Wilson
loop operators
\cite{MalWilson,PolRyc,GroDru,Berenstein:1998ij,Semenoff}. Here one is aided
by the fact that $\lambda=g_{YM}^2 N$ is precisely the semi-classical
expansion parameter for the loop equations.

In this paper we study AdS/CFT correspondence in the opposite limit
viz. $\lambda \rightarrow 0$. In this limit the gauge theory can be
analyzed perturbatively but the string theory is more difficult and
the standard semi-classical expansion of strings when
$R^2=\alpha'\sqrt\lambda$ is large, breaks down. The problem is
conceptually similar to the treatment of gauge theories on the lattice
\cite{Wilson:1974sk,Polyakov:rs}
where the gluon picture breaks down at strong coupling. 
The zeroth order solution of the lattice gauge theory is obtained by
exactly diagonalizing the kinetic (square of the electric field) term
of the Hamiltonian. Gauge invariance (Gauss' law) then connects the
individual string bits on the links of the lattice into long strings
which can connect charged sources.  The potential energy (magnetic
term) is treated perturbatively and it corrects the energy and wave
function of the string state of the zeroth order Hamiltonian
\cite{KogSus}.

We adopt a similar method, for the strongly coupled world sheet string
theory, when $\lambda \rightarrow 0$. We assume that $N$ is large
enough so that string loops are suppressed and the first quantized
picture is an adequate starting point. We work in the light cone gauge
which leaves only global reparametrizations of the string as a
residual symmetry. We consider the parameter space of the gauge-fixedstring
as a discrete lattice consisting of $M$ string bits.  At $\lambda =
0$, the string bits are non-interacting. On each bit we have a Hilbert
space which turns out to consist of the states of a super-particle in AdS$_5
\times S^5$. The total wave function of the string is a cyclically
symmetric direct product of states carried by the bits. We explicitly
present these states.

Bit strings have been discussed earlier in the context of strings in
flat space\cite{Giles:1977mp} and more recently in the pp-wave
background \cite{BMN,Kristjansen:2002bb,G7,Zhou,HV}. A problem similar
to the one treated here is that of bosonic strings in AdS$_5$ in the
small radius limit, which has been addressed in \cite{Karch:2002vn},
using somewhat different methods. A sensible perturbative treatment
is unlikely to exist in this case since fermionic contributions turn
out to be crucial to give finite results in our calculations.  The
idea of string bits has also appeared in 't Hooft's paper
\cite{tHooft} on planar diagrams.

The organization of this paper is as follows. In the next section we
will review the quantization of Green-Schwarz world-sheet type IIB
string theory in the light-cone gauge. This gauge-fixing uses Poincare
co-ordinates. In section 3 we will discuss a limit of this theory in
which the AdS radius vanishes. In this limit the theory reduces to a
system of non-interacting bits. We make this more explicit by
introducing discrete bits of string. Each bit behaves like a
superparticle moving in the AdS$_5 \times$ S$^5$ background, 
carrying ``appropriate representations'' of the superconformal 
group\footnote{\label{fn:1a}In the light cone gauge, some of the 
generators of the superconformal algebra, which involve $\tilde x^-$, the 
non-zero mode of $x^-$, do not have well-defined action on a single bit and 
one needs to consider at least two bits to define their action.  A single
bit carries a representation of the superconformal algebra only in the
naive sense that $\tilde x^-$, which is defined in terms of gradients
of the independent coordinates or momenta, is taken to be zero by
fiat. We will discuss the action of these generators on multi-bit
states in Appendix D.}. In the light-cone gauge, the operator that
gives evolution in light-cone time has continuous eigenvalues. This
basis is not convenient for comparison with the gauge theory at the
boundary. For this purpose, it is desirable to work with the basis
frequently used in describing the UIR's of the superconformal group
PSU$(2,2|4)$. We identify the operators in the light-cone gauge which
have this basis as eigenstates\footnote{\label{fn:1b} One can
encounter continuous or discrete eigenvalues in a given physical
system depending on which operator is being diagonalized, e.g. in the
case of the two-dimensional rotor we have continuous eigenvalues of
linear momentum but discrete eigenvalues of angular momentum.}. In
section 4 we discuss the states of a single bit in such a basis.  In
section 5 we discuss corrections to the multi-bit spectrum due to a
small but non-vanishing value of the AdS radius. A non-vanishing value
of the radius introduces bit-bit interactions which modifies the
spectrum.  We compute the leading order corrections to the simplest,
but potentially the most singular, multi-bit states. We find that
potential divergences from the boundary $z=0$ cancel between fermionic
and bosonic contributions, leaving only finite corrections 
to the anomalous dimensions. In section 6 we make
preliminary attempts at constructing string states out of multi-bit
states. There are difficulties in doing this, which we discuss in
this section. A naive continuum limit, which is arranged so as to give
finite energy states, seems to be in conflict with unitarity 
bounds\footnote{We thank Shiraz Minwalla for discussions on this point.} 
on the representations of the superconformal group.
This motivates us to consider the alternative possibility that 
discretization is essential. In this case one must include
string states made of varying number of bits. This is because splitting
and joining of strings with a given number of bits will produce strings with
lower as well as higher number of bits. However, we find that
the string spectrum made this way does not agree with that of the
gauge theory operators. We discuss this in this detail in this
section.  We end with some discussion and concluding remarks in
section 7. Appendix A contains the continuum expressions for
superconformal generators as realized in our string theory, and
Appendix B the discrete version. Appendix C contains details of
perturbation calculation and the cancellation of divergences. Appendix
D contains a discussion on stringy features of the superconformal
generators and their action on multi-bit states.

In this paper, we will work in the $N \to \infty$ limit and
ignore string-string interactions.

\section{Type IIB strings in AdS background in the light-cone gauge} 

In this section we will review type IIB Green-Schwarz superstring in
the AdS$_5 \times$ S$^5$ background geometry in the light-cone gauge
\cite{MT, MTT}. This will also serve to set our notations and
conventions. We will use Poincare co-ordinates since the
discussion of light-cone gauge fixing is simplest in these
co-ordinates.  Also, in these co-ordinates comparison with the
boundary gauge theory is the most direct. We will not address global
issues arising out of the use of a Poincare patch in the present work.

The AdS$_5 \times$ S$^5$ metric in Poincare co-ordinates is
\bea
ds^2={{\rm R}^2 \over z^2}(\eta_{\mu\nu} dx^\mu dx^\nu+dz^adz^a),
\label{twoone}
\eea 
where $\eta_{\mu\nu}=(-, +, +, +)$ is the Minkowski metric and the six
co-ordinates $z^a, a=1,2, \cdots, 6$ parametrize the S$^5$ and the AdS
``radial'' direction.  The transformation between the ``Cartesian''
$z^a$ and the more customary $z, \Omega_5$ is given by
\[
z^a= z \n^a, \quad \n^a\n^a=1, 
\]
where the six-dimensional unit vector $\n^a$ represents $S^5$ angles
$\Omega_5$. The $z\to 0$ limit is called the AdS boundary whereas $z\to
\infty$ is called the Poincare horizon. 

The AdS radius $R$ is related to the 't Hooft coupling
$\lambda$ and the string scale $\alpha'$ by the standard
expression 
\be 
{\rm R}^2=\alpha'\sqrt\lambda, \quad \lambda=g_{\rm YM}^2 N
\label{coupling}
\ee
Note that in the world-sheet action the AdS radius $R$ naturally
appears in the combination ${\rm R}^2/\alpha'=\sqrt\lambda$.
This is the  effective tension of the string moving in
AdS$_5 \times$ S$^5$. For later convenience we introduce the
parameter
\be
{\rm T}= \frac{\sqrt \lambda}{2\pi}
\label{def-t}
\ee

In addition to the bosonic co-ordinates, appearing in (\ref{twoone}),
the world-sheet superstring action has two left Majorana-Weyl
spinors. In the light-cone gauge, the $\kappa$-symmetry of the
Green-Schwarz action can be used to gauge away half of the fermionic
degrees of freedom of each of these two spinors \cite{MT}. The remaining $16$
physical fermionic degrees of freedom can be rearranged into two
fundamental representations of the R-symmetry group SU$(4)$,
$\theta^i, \eta^i, i=1, \cdots, 4$ and their conjugates, $\theta_i,
\eta_i$ . The $\theta^i$ are related to the $8$ linear Poincare
supersymmetries while the $\eta^i$ are related to the $8$ non-linear
``conformal'' supersymmetries of the $psu(2,2|4)$ superconformal
algebra. In fact, $\theta$ appears at most quadratically in the 
generators of this algebra, while there are also terms quartic in $\eta$ 
in these generators. These latter terms reflect the presence of curvature 
and the five form flux.

On the bosonic side, it turns out \cite{MTT} that the usual
light-cone gauge choice
\[
x^+ \equiv {x^3+x^0 \over \sqrt{2}}=p^+ \tau,
\] 
where $\tau$ is the world-sheet time co-ordinate, cannot be combined
with a conformal gauge on the world-sheet metric since that is not
consistent with equations of motion of the system. A modified gauge
choice for the world-sheet metric, $g_{\alpha\beta}$, which is
consistent with the gauge choice on $x^+$, is \cite{MTT}
\be
\sqrt{-g} g^{\tau\tau}={-1 \over \sqrt{-g} g^{\sigma\sigma}}=-{2\pi
z^2 \over \sqrt{\lambda}},
\label{gtt-gauge}
\ee
where $\sigma$ is the world-sheet co-ordinate running along the
string\footnote{This gauge choice is potentially singular near the
boundary, $z \rightarrow 0$. However, as we shall see later in this
paper, physical quantities seem to be regular, so presumably this is
just a gauge artifact. Note that $\lambda\to 0$ is not
a problem since the Hamiltonian \eq{twotwo} is
well-defined in this limit.}.

Before gauge fixing, the Green-Schwarz type IIB superstring moving in
AdS$_5 \times$ S$^5$ background manifestly has the global isometries
of the supergroup PSU$(2,2|4)$. The choice of a light-cone gauge
destroys manifest symmetry, but one can still derive expressions for
the generators of the symmetry algebra in terms of the light-cone
bosonic and fermionic degrees of freedom of the superstring \cite{MTT}. 
For example, the ``Hamiltonian'' which generates evolution in the
light-cone time variable $x^+$, and is one of the generators of the
superalgebra $psu(2,2|4)$, is given by $-\int d\sigma \ {\cal
P}^-(\sigma)$, where
\bea
{\cal P}^-\!\! &=&
\!\!  -{1 \over 2 p^+}\biggl(2 p^x p^{\bar x}- \del_z^2+
{1 \over z^2} \big[ l^i_jl^j_i +
4\eta_i {l^i}_j \eta^j+(\eta_i\eta^i -2)^2-\frac14\big]\biggr)-
{{\rm T}^2 \over 2p^+ z^4}(2x'\x'+(z_a')^2) \nn \\
&& +{{\rm T} \over z^2 p^+} \eta^i\rho^a_{ij}\n^a(\theta'^j-
i{\sqrt{2} \over z} \eta^j x')
+{{\rm T}\over z^2 p^+} \eta_i{\rho^a}^{ij}\n^a
(\theta'_j+i{\sqrt{2} \over z} \eta_j {\bar x}'), 
\label{twotwo}
\eea
Here, T is given by \eq{def-t}. The notation is as in Appendix A, in
particular ${l^i}_j$ is as in \eq{A19}.  In this appendix, for the
convenience of the reader, we have also given the light-cone gauge
expressions for the basic generators of $psu(2,2|4)$ algebra, together
with their (anti)commutation relations.  This algebra, and the
light-cone gauge realization of it, will play a central role in what
follows.

\section{String bits and the limit of zero radius}

We begin by noting that due to the gauge choice \eq{gtt-gauge} the AdS
radius R enters the expression in (\ref{twotwo}), through T, only in
the terms that have $\sigma$-derivatives. This is, in fact, the case
with all the generators\footnote{except for generators mentioned in
footnote \eq{fn:1a} on page \pageref{fn:1a} which we discuss in
Appendix D.}, as can be seen from the expressions given in 
Appendix A.
As a consequence of this, in the limit R $\rightarrow 0$ all the
$\sigma$-derivative terms disappear from these generators and they
become pointwise local in $\sigma$. That is, the string ``breaks up''
into independent bits. Since the finite T terms are apparently
singular at the boundary $z \rightarrow 0$, one might worry whether it
makes any sense to think of a non-zero radius in (\ref{twotwo}) as a
perturbation. Later in this paper we will see by explicit calculations
that perturbation expansion in T is finite. With this post facto
justification, we will just go ahead with the naive zero radius limit.

To make the bit picture manifest, let us discretize $\sigma$ into a
lattice of $M$ points with a lattice spacing $\epsilon = l/M$, where
$l$ is the total length of the string. As usual, we need to rescale
continuum variables by powers of $\epsilon$ to get the lattice
variables. Thus, the fact that the total light cone momentum
$P^+=\int_0^l d\sigma \ p^+_{\rm cont.}  =lp^+_{\rm cont.}$ must be
equated to $M p^+_{\rm latt.}$ implies that the momentum of each bit
must be $p^+_{\rm latt.} = \epsilon p^+_{cont.}$. We will permit
ourselves the abuse of notation $p^+_{\rm latt.}  =p^+$ when we are
discussing bit variables.  Similarly, the momentum densities (at any
point $\sigma = n \epsilon$ along the string) in the directions $(x^i,
z^a), i=1,2; a=1,...,6,$ are related to the discrete momenta of the
$n$th bit by relations such as $p^i_n = \epsilon p^i(\sigma)$.  This
implies that (a) the total momenta reduce to simple sums, e.g.
$P^i=\sum_1^M p^i_n$, and (b) since we should not rescale the $x$'s,
so that the continuum canonical commutation relation, $[x^i(\sigma),
p_j(\sigma')]=i\delta^i_j \delta(\sigma-\sigma')$, translates to the
discrete expression $[x^i_n, p_{m,j}]=i\delta^i_j \delta_{mn}$, which
is free of $\epsilon$.  Here we have used $\delta(\sigma-\sigma')=
(1/\epsilon) \delta_{mn}$.  A similar rescaling of the superpartners,
$\sqrt{\epsilon}\theta \rightarrow \theta, \sqrt{\epsilon}\eta
\rightarrow \eta$ removes the lattice spacing from their
anti-commutation relations. The above rescalings are summarized in
\eq{rescalings}.
 
In terms of these discrete bit variables, the T-independent part of
the Hamiltonian, is given by
\be
P_0^- \equiv \sum^M_{n=1} p^-_{n}=-\sum^M_{n=1}
{1 \over 2 p^+} \biggl(2p^x_np_n^{\bar x}-
\del_{z_n}^2 +{1 \over z_n^2}\biggl[(l^i_{n,j} l^j_{n,i})+4\eta_{n,i} 
{l^{i}}_{n,j} \eta^j_n+ (\eta_{n,i}\eta^i_n -2)^2-\frac14
\biggr]\biggr)
\label{threeone}
\ee
The subscript on $P_0^-$ denotes that T has been set to zero. 
We see that the lattice spacing has completely disappeared from the
expression on the right hand side of (\ref{threeone}). In fact, this
continues to be so even when finite T parts are included. This
property of the Hamiltonian is actually shared by all the generators
of the superconformal algebra. 

In the limit of vanishing AdS radius we thus see that the string ``falls
apart'' into non-interacting discrete bits.  The first step in the 
problem of solving for the spectrum of states of the free
(i.e. $\lambda=0$) superstring then reduces to the problem of solving
for the spectrum of a single string bit\footnote{Provided that
generators which involve $\tilde x^-$, such as $K^x$, which a priori
do not have any well-defined action on a single bit Hilbert space, act
appropriately on these tensor product states. See footnote
\eq{fn:1a}.}. It is to this problem that we will now turn our
attention. In Appendix B we have listed discretized expressions for
all the generators, including the finite T parts. We have also listed
there the canonical commutation relations satisfied by the bits. We
will need these expressions in the discussions that follow.

\section{States of a single bit}

In thinking of the spectrum of states of a single bit, the first
difficulty that we face is that the spectrum of the single-bit
Hamiltonian $-p^-_{n}$ is continuous. Clearly the eigenstates of
$-p^-_{n}$ are not a good starting point for building free superstring
states since these states cannot be directly compared with the
gauge-invariant operators of the boundary theory. This is because the
latter have definite integer or half-integer dimensions in the limit
of vanishing 't Hooft coupling $\lambda$. It is, however, easy to
resolve this problem. The unitary irreducible representations (see,
e.g. \cite{Mack,Dobrev,GM,KRN,SM} of the global symmetry group SO$(4,
2)$ of AdS$_5$ are well-known to be labeled by three numbers
corresponding to the representations of the maximal compact subgroup
SO(2) $\times$ SO(4)= SO(2) $\times$ SU(2) $\times$ SU(2). Here SO(2)
denotes rotation in the $(-1,0)$ directions (generated by $S_{-1,0}$
of \eq{A1}) while SO(4) denotes rotation in the remaining four
directions (see \eq{A1},\eq{hyperb} for notation). These quantum
numbers are denoted as $(E_0, J_1, J_2)$ (\cite{GM}, see also
\cite{ZFF}), where $E_0$ physically means the energy, conjugate to
time translations in global co-ordinates, and $J_1$ and $J_2$ denote
the angular momentum representations associated with the two SU(2)
factors in the SO(4). (see equations in section A.2
for more detail). Equations \eq{e0} and \eq{so4} 
express these generators in terms of the generators of the conformal algebra in
$(3+1)$-dimensional Minkowski space. It turns out from these
considerations (see Sec 2.2 of \cite{SM}) that the Euclidean dimension
operator of the conformal algebra (equivalently the dimension in the
boundary gauge theory) corresponds to $E_0$.

Clearly, the spectrum should be arranged in terms of simultaneous
eigenfunctions of $(E_0, J_1, J_2)$, or equivalently, $(H_+, H_-,
J_1-J_2)$, where $H_\pm \equiv E_0 \mp (J_1 +J_2)$ (on a suitable
highest weight state, see \eq{j1-j2}). The latter basis is
particularly convenient because these operators have 
simple expressions in the light-cone frame, viz.
\bea
H_+=-P^-+K^+, \quad H_-=-K^-+P^+, \quad J_1-J_2=J^{x {\bar x}}.
\label{threetwo}
\eea

Thus, we should find simultaneous eigenstates of these operators
\footnote{Similar operators have also been considered
in \cite{Metsaev:2002vr}.}
(and {\em not} of $-P^-$) in order to determine which UIR's of SO$(4,
2)$ appear. Representations of the full superalgebra,
which includes additional bosonic generators (of the $SU(4)$
R-symmetry), are constructed in our case by acting on the
abovementioned eigenstates with supersymmetry ``raising operators''
\eq{raising} (cf. \cite{GM,Dobrev,SM}). 

\subsection{The spectrum of $h_+$}

In this section we will determine the spectrum of  the single-bit
operator $h_{+,n}$ where 
\be
H_+= \sum_{n=1}^M h_{+,n}
\label{h-def}
\ee
in the limit $\lambda=0$. In a later section we will determine the
perturbative correction to the spectrum of $H_+$ at small
$\lambda$. Since $H_+ = E_0 - (J_1+J_2)$ on a highest weight state, as
mentioned above, and since $J_1-J_2$, being an integer or a
half-integer, does not receive perturbative corrections, the
perturbative correction to the spectrum of $E_0$ is the same as that
of $H_+$. Henceforth, we will sometimes use the word ``Hamiltonian'' 
for the operator $H_+$.

We will see that it is easy to determine $J^{x \bar x}$ on the
eigenfunctions of (\ref{h-def}). In order to determine the triple
$(E_0, J_1, J_2)$, we also need to evaluate $H_-$; this is not easy
because it involves $\tilde x^-$ and is hence non-local even at
$\lambda=0$. However, we will be able to measure $H_-$ indirectly on
states of our interest
\footnote{\label{anom} We stress that $J_1, J_2$ are both
(half-)integers and are immune from perturbative corrections.  Since
$H_- - H_+ = 2 (J_1 + J_2)$, it is enough to consider perturbative
correction to $H_+$ which, as argued earlier, represents also the
perturbative correction to $E_0$.}

The expression for $h_+$, as defined in \eq{h-def},
can be obtained from the equations (\ref{B2}) 
and (\ref{B7}) given in Appendix B:
\be
h_+ =-\frac{\del_x \del_{\bar x}}{p^+} + p^+ x \bar x
-{\del_z^2 \over 2p^+}+{p^+ \over 2}z^2 
+{1 \over 2p^+z^2} \big[({l^i}_j)^2+4\eta_i {l^i}_j \eta^j+
(\eta_i\eta^i-2)^2-{1 \over 4} \big]
\label{threefive}
\ee
Here and in the rest of this section we will drop the bit index $n$
from single bit operators to simplify the notation. Note that the
fermionic terms in the above come entirely from the $-p^-$ part of
$h_+$.

\subsubsection{\label{ho}States with no fermions=
eight bosonic oscillators}

To find the purely bosonic states, it is more instructive first to
transform from the ``polar coordinates'' $(z, \n_a)$ used in
\eq{threefive}, to ``Cartesian coordinates'' $z_a= z \n_a, a=1,...,6$.
The Hamiltonian appropriate to the Cartesian coordinates, denoted
by $\tilde h_+$, is given by
\bea
\tilde h_+ &&=z^{-5/2} h_+  z^{5/2}
\nn\\
&&=
{p^x p^{\bar x} \over p^+}+ p^+ x {\bar x}+
{(p^a)^2 \over 2p^+}+{p^+ \over 2}(z^a)^2 +{1 \over 2p^+z^2}
(4\eta_i {l^i}_j \eta^j+(\eta_i\eta^i)^2-4\eta_i\eta^i).
\label{threethree}
\eea
Here $h_+$ refers to \eq{threefive}. The eigenfunctions
of \eq{threethree}, $\tilde \psi$, are related to eigenfunctions
$\psi$, by  $\tilde \psi = z^{-5/2} \psi$. Here $p^x
= - i\del_{\bar x}, p^{\bar x}= -i \del_x,
p^a = -i \del/\del z_a$.

On states that do not contain any fermions, the coefficient of $1/z^2$
in (\ref{threethree}) vanishes. The rest of the terms represent
{\em eight harmonic oscillators} corresponding to the eight transverse
bosonic degrees of freedom of the light-cone gauge-fixed string.  The
spectrum of $\tilde h_+$ (hence of $h_+$) is obviously discrete
(in fact, integer\footnote{\label{integer}
Although each of the triple $(E_0, J_1, J_2)$ can be an
integer or half integer, it turns out that $H_+
\equiv E_0 - (J_1+ J_2)$ is always an integer.}), and is given by
\be
E= \sum_{\iota=1}^8 (n_\iota  + 1/2), \quad n_\iota=0,1,...
\label{ho-spectrum}
\ee
The length scale of the harmonic oscillators is set by $p^+$, e.g. $\langle
x\bar x \rangle = \langle (z^a)^2 \rangle= 1/p^+$, although the
spectrum is free of $p^+$. The creation and annihilation operators are
defined in  Appendix B (see \eq{B1a}).

Note that the existence of the eight bosonic oscillators ties up with
the fact that there are precisely eight fermionic oscillators $\eta_i,
\theta_i, i=1,...,4$, as we expect from supersymmetry.

\subsubsection{General wavefunctions}

Let us now consider the more general case when fermionic degrees of
freedom are excited. We will first write down the result.

The spectrum  of $h_+$ (\eq{threefive}) is given by
(see footnote \eq{integer})
\be
e_+ = \alpha + 2 r + s + \bar s + 2,\quad  \alpha, r, s, \bar s \in 
{\mathbf Z}
\label{gen-spectrum}
\ee
where the corresponding eigenfunction is given by 
\be
\ket \Psi = 
\ket{\eta}\ket{\theta}\times 
\Upsilon^\alpha_{l;s,\bar s,r}(x,\bar x, z)
\times \ket{l}
\label{gen-wave}
\ee
Here $\ket{\eta}\ket{\theta}$ denotes the  part of the wavefunction
involving fermionic coordinates. $\ket{\eta}$ is  any 
polynomial in the four variables $\eta_i, 
i=1,\ldots, 4$, understood as acting on the fermionic ground state
$\ket{0}$. There are 16 such linearly independent 
polynomials: $1, 
\ \eta_i,\ \eta_i \eta_j, \ldots, 
\eta_1\eta_2\eta_3\eta_4 $. Similarly $\ket{\theta}$ denotes 
any polynomial in $\theta_i, i=1,\ldots, 4$. Recall that 
$\{\theta_i, \theta^j\}=
\{\eta_i, \eta^j\} = \delta^j_i, \; \eta^i\ket{0}=0
=\theta^i\ket{0}$. 

The various parts of the wavefunction depending on the bosonic
coordinates are as follows.
\be
\Upsilon^\alpha_{l;s,\bar s,r}(x,\bar x, z) := h_{s,\bar s}
(x,\bar x)\ \psi^\alpha_{p^+,r}(z)  
\ee
Here $h_{s,\bar s}(x,\bar x)$ is the standard harmonic oscillator
wavefunction of a pair of complex coordinates (see
\eq{threethree} and Appendix B) at level $s, \bar s$ respectively, and
\be
\psi^\alpha_{r}(z) = e^{-p^+ z^2/2}
z^{\alpha+{1\over2}} 
L_r^\alpha(p^+ z^2) \sqrt{\frac{2.r!}{(r+\alpha)!}},
\label{psi-alpha}
\ee
where $L_r^\alpha$ are generalized Laguerre functions.
$\alpha$ is an integer defined below and
$\int_{-\infty}^\infty dz\psi^\alpha_r(z)^2=1 $ for $p^+=1$.

The wavefunction $\ket{l}$ depends on the $S^5$ angles,
equivalently on the $\n_a, a=1,..6$, and is given by
the spherical harmonic
\[
\ket{l}= C_{a_1...a_l} \n^{a_1}...\n^{a_l}
\]
where $C_{a_1...a_l}$ is a symmetric, traceless, rank $l$ tensor.
$\ket{l}$ satisfies
$ l^{ab}l^{ba} \ket{l} = l(l+4) \ket{l} $ ($l^{ab}$ is 
defined in \eq{A19}).

\noindent\underbar{Definition of $\alpha$}

The integer $\alpha$ appearing in \eq{gen-spectrum} and
\eq{gen-wave} and onwards, depends on the $S^5$ angular
momentum $l$ and the fermion wavefunction $\ket{\eta}$,
as follows \cite{Met}.

\gap2

\noindent 
Case $l=0$: $\alpha= | 2 - F_\eta |$ where $F_\eta$ is the number of
$\eta$'s in the wavefunction $\ket{\eta}$.

\gap2

\noindent
Case $l\not= 0$:  
\begin{enumerate}
\item $F_\eta=0$ or $F_\eta=4$: $\alpha= l+2$.
\item $F_\eta=1$: the canonical choices for
the wavefunction $\ket{\eta}$, viz. $\eta_i, i=1,...,4$
do not diagonalize $h_+$. The operator $h_+$ splits the
4-dimensional space ${\V}_1= {\rm Span}\{\eta_i \} \equiv$
complex linear combinations of $\{\eta_i\}$, into two
2-dimensional spaces given by $P_1 {\V}_1$ and $ Q_1 {\V}_1$,
where $P_1, Q_1$ are orthogonal projection operators,
\be
(P_1)^i_j= \frac{l+4}{2l + 4}
\biggl(\delta^i_j - \frac2{l+4} l^i_j\biggr),
\ (Q_1)^i_j= \frac{l}{2l + 4}
\biggl( \delta^i_j + \frac2{l} l^i_j\biggr)
\label{p1}
\ee
If $\ket{\eta}$ belongs to $P_1 {\V}_1$, then $\alpha=l+1$,
if $\ket{\eta}$ belongs to $Q_1 {\V}_1$, then  $\alpha=l+3$.
\item
$F_\eta=3$: Once again the four canonical 
basis vectors $\eta_i \eta_j \eta_k$
do not diagonalize $h_+$. The operator $h_+$ splits the
4-dimensional space 
${\V}_3= {\rm Span}\{\eta_i \eta_j \eta_k\}$ into two
2-dimensional spaces given by 
$P_3 {\V}_3$ and $ Q_3 {\V}_3$, where $P_3, Q_3$
are orthogonal projection operators,
\bea
&& (P_3)^{ijk}_{mnp}= \frac{l+4}{2l + 4}\biggl(
\delta^i_{[m}\delta^j_n\delta^k_{p]} - \frac6{l+4} 
\delta^i_{[m}\delta^j_n l^k_{p]}
\biggr),
\nn\\ 
&& (Q_3)^{ijk}_{mnp}= \frac{l}{2l + 4}\biggl(
\delta^i_{[m}\delta^j_n\delta^k_{p]}+ \frac6{l} 
\delta^i_{[m}\delta^j_n l^k_{p]}\biggr)
\label{p3}
\eea
Here $[\ ]$ on the subscripts denote antisymmetrization
with weight unity.
If $\ket{\eta}$ belongs to $P_3 {\V}_3$, then $\alpha=l+1$,
if $\ket{\eta}$ belongs to $Q_3 {\V}_3$, then  $\alpha=l+3$.  
\item
$F_\eta=2$: Here the operator $h_+$ splits the
6-dimensional space ${\V}_2= {\rm Span}\{\eta_i \eta_j\}$ into three
2-dimensional spaces given by $P_2 {\V}_2, Q_2 {\V}_2 $ and
$R_2 {\V}_2$, where $P_2, Q_2, R_2$
are orthogonal projection operators,
\bea
(P_2)^{ij}_{mn} &&=\frac{l+4}{4(l+1)} \biggl(
\delta^i_{[m}\delta^j_{n]} -  
\frac{4(l+3)}{(l+2)(l+4)}\delta^i_{[m} l^j_{n]} +
\frac{4}{(l+2)(l+4)}l^i_{[m} l^j_{n]} \biggr)
\nn\\
(Q_2)^{ij}_{mn} &&=  \frac{l(l+4)}{2 (l+1)(l+3)}\biggl(
\delta^i_{[m}\delta^j_{n]}
+ \frac{8}{l(l+4)}\delta^i_{[m} l^j_{n]}
- \frac{4}{l(l+4)}l^i_{[m} l^j_{n]} \biggr)
\nn\\
(R_2)^{ij}_{mn} &&= \frac{l}{4(l+3)}\biggl(
\delta^i_{[m}\delta^j_{n]}
 +\frac{4(l+1)}{l(l+2)}\delta^i_{[m} l^j_{n]} +
\frac{4}{l(l+2)}l^i_{[m} l^j_{n]} \biggr)
\label{p2}
\eea
If $\ket{\eta}$ belongs to $P_2 {\V}_2$, then $\alpha=l$,
if $\ket{\eta}$ belongs to $Q_2 {\V}_2$, then  $\alpha=l+2$,
if $\ket{\eta}$ belongs to $R_2 {\V}_2$, then  $\alpha=l+4$.
\end{enumerate}

\noindent\underbar{The value of $J^{x \bar x}$}

\eq{gen-wave} is also an eigenfunction of $J^{x \bar x}$
(see \eq{A11}), with eigenvalue
\[
J^{x \bar x}= - s + \bar s + \frac12 (F_\eta - F_\theta)
\]
 
\noindent\underbar{A note about normalizability}

Note that in \eq{gen-wave} we have considered only {\em normalizable}
solutions 
\footnote{In the light-cone gauge, the time evolution of the states of
a bit is governed by a Schrodinger type evolution equation in flat
space with a potential. The appropriate norm in this case is the usual
Schrodinger norm.}. This is appropriate for constructing the space of
states of the superstring. Alternatively, this can be seen from
requiring that the solution be regular at $z
\rightarrow \infty$. This condition uniquely picks the normalizable
solution because of the harmonic oscillator potential piece in
(\ref{threefive}).

\subsubsection{\label{1bit-comparison}Summary of this subsection}

The spectrum of states for a bit that we have obtained here is
identical to that of the type IIB supergravity multiplet
\cite{MTT,Tseytlin:2002gz}. For instance, the 20-plet of states
\be
\ket{A}= A^{ij,kl}\eta_i \eta_j \theta_k \theta_l \ket{0}\times 
\Upsilon^0_{0;0,0,0}(p^+,x,\bar x, z)\ \ket{l=0},
\label{basic-state}
\ee
where the tensor $A^{ij,kl}$ belongs to the ${\bf 20'}$, i.e.
the Dynkin label $(0,2,0)$ of SU(4),
corresponds to the part of the supergravity multiplet
characterized by $(2,0,0| {\mathbf{20'}})$. The fact that
the $so(4,2)$ quantum numbers $(E_0,J_1,J_2)$ are
given by $(2,0,0)$ follows from the fact that
$h_+=2, j^{x\bar x}=0$ and, as we discuss in Appendix D,
$h_-=2$. 

Note that since the states of a single bit just constitute the
supergravity multiplet these states are all part of a short
multiplet. This is consistent with the non-renormalization theorems,
at least in the limit in which string interactions are ignored, since
a single bit cannot receive corrections due to bit-bit interactions.

\section{Multi-bit spectrum}

The first step in the construction of the states of a free string from
the states of a bit is to construct multi-bit states. To do this one
needs to note the following points:

\begin{enumerate}

\item The phase space of the string is given by the canonically
conjugate pairs in \eq{funda-canon}. This shows that $p^+$ does not
depend on $\sigma$; consequently all bit wavefunctions in the
multi-bit state must have the same value of $p^+$. The final multi-bit
wavefunction for $m$ bits contains a function of $p^+$ in addition to
the plane wave ${\rm exp}(imp^+x^-_0)$. This is because multi-bit
wavefunctions do not have a definite value of $p^+$ in the basis in
which $H_-$ is diagonal.

\item Physical string states must satisfy the constraint coming from the
residual global reparametrization symmetry in the light-cone gauge. In
the discrete version what this means is that to get a physical state of a
string from a multi-bit state with $m$ bits, one should cyclically symmetrize a
tensor product state of $m$ bits. Thus, we should only consider 
multi-bit states of the following form:
\be
\ket{p^+; \psi_1, \Psi_2, \cdots}
= \sum_{\sigma\in P}
\ket{p^+; \psi_{\sigma(1)}} \otimes 
\ket{p^+; \psi_{\sigma(2)}} \otimes \cdots \otimes 
\ket{p^+; \psi_{\sigma(m)}}.
\label{fourone}
\ee
Here $P$ is the group of cyclic permutations of $m$ variables.  We
have inserted the value of $p^+$ as an explicit reminder of point (1)
above. Henceforth, in future references to \eq{fourone} the $p^+$
factors will not be explicitly written.

\end{enumerate}

\noindent In the rest of this section we will first discuss the multi-bit
states at zero radius and then compute corrections to the spectrum of these
states due to a non-zero, but small, value of the radius.

\subsection{Multi-bit spectrum at $\lambda=0$}

For the state \eq{fourone}, the value of $H_+$ at $\lambda=0$ is
clearly given by
\be
E_{+,0} = \sum_{n=1}^m e_{+,n}
\label{multi-spectrum-0}
\ee
where $ e_{+,n}$ denotes the value \eq{gen-spectrum} for
the state  $\ket{\psi_n}$.
As an example, we can construct a state \eq{fourone}, where we take
each wavefunction to be of the type given by \eq{basic-state}.
Explicitly, the state will look like
\be
\ket{\Psi_0}= {\cal A}_{i_1, j_1,...,k_m, l_m}\prod_{n=1}^m
\eta_{i_n}\eta_{j_n} \theta_{k_n} \theta_{l_n} \ket{0}
\Upsilon^0_{0;0,0,0}(x_n, \bar x_n, z_n)\ket{l=0}
\label{basic-multi-bit}
\ee
Here we have allowed for a general polarization ${\cal A}$ rather than
a simple cyclically symmetrized product of individual polarizations
$A^{(n)}_{i_n, j_n; k_n, l_n}$.
If we take the polarization ${\cal A}$ to correspond to a traceless
symmetric product of the individual polarizations $A^{(n)}$'s, then
the state becomes 
\be
\ket{\Psi_{\rm sym}}= {\cal A}^{\rm sym}_{i_1, j_1,...,k_m, l_m} 
\prod_{n=1}^m
\eta_{i_n}\eta_{j_n} \theta_{k_n} \theta_{l_n} \ket{0}
\Upsilon^0_{0;0,0,0}(x_n, \bar x_n, z_n)\ket{l=0}
\label{sym-wave}
\ee
Under $psu(2,2|4)$, this state transforms as 
$ (2m,0,0| \mathbf{0,2m,0})$ which is a BPS state made out of $m$ bits
(this can be seen by noting that, in the notation of Gunaydin et al \cite{GM},
this corresponds to the multiplet based on the Fock space vacuum $\ket{0}$ 
with $p=2m$ oscillators).

\subsection{Finite radius correction to multi-bit spectrum}

In this section we will compute corrections to the free spectrum, 
\eq{multi-spectrum-0}, of multi-bit states due to bit-bit 
interactions introduced by turning on a small
T$= \sqrt \lambda/(2\pi)$ (see \eq{def-t}). Let us write
\[
H_+ = H_+^0 + {\rm T} V_1 + {\rm T}^2 V_2
\]
where $H_+^0, V_1, V_2$ are defined in
\eq{h0-v1-v2}. 
The leading correction to \eq{multi-spectrum-0} is given
by T$^2\Delta E_+$, where
\be
\Delta E_+ =
\langle \Psi_0 | V_2 | \Psi_0 \rangle -
 {\sum_{int}}' 
\frac{ \langle |\Psi_0 | V_1 | \Psi_{int} 
\rangle|^2}{E_{+,int}-E_{+,0}}
\label{delta-e}
\ee
In Appendix C we present a calculation of $\Delta E_+$
for the state $\ket{\Psi_0}$ given in \eq{basic-multi-bit}.
Here we present the main points. 

\subsubsection{Potential divergence from the region $z\to 0$}

Consider the $V_2$ term in \eq{delta-e}. Let us look at the term
involving $(x_{n+1} - x_n)^2/ z_n^2$ in $V_2$ (see
\eq{h0-v1-v2}). The numerator and the denominator separate out in the
expectation value: the first gives a simple factor of $1/p^+$ and the
second an integral of the kind
\[
\int_0^\infty \frac{dz_n}{z_n^4} |\psi^{\alpha=0}_{r=0}(z_n)|^2
\] 
Using $\psi^0_0 (z) \sim \sqrt z$ near $z=0$ (see \eq{psi-alpha}) the
above integral, summed over $M$ bits, gives a leading divergence
\be
M \int_{z_{min}}^\infty \frac{dz}{z^3} \sim M\ \frac{1}{z_{min}^2}
\label{v2-div}
\ee
Here we have put an infra-red cut-off $z_{min}$ near the boundary 
$z=0$. 

As for the $V_1$ contribution in \eq{delta-e} (see
Appendix C for details), the  matrix element 
$ \langle \Psi_0 | V_1 | \Psi_{int} \rangle$ has
finite $z$-integrals of the type
\be
\sum_n
\int_0^\infty 
\frac{dz_n}{z_n^3} \psi^{\alpha=0}_{r=0}(z_n) \psi^{\alpha=2}_{r}
(z_n) = M C_1(r)
\label{c1-r}
\ee
where $C_1(r)$ also includes the finite angular and fermion matrix
elements. The $z$-integral is finite (esp. at $z\to 0$), being $\sim
\int dz z^{1/2 + 7/2}/z^3$, using the fact that for $z\to 0$,
$\psi^\alpha_r (z) \sim z^{\alpha+1/2}$ (see
\eq{psi-alpha}).  The initial state has $\alpha=0, r=0$.  The reason
for the appearance of $\alpha=2$ in the intermediate state is that
$V_1$ contains (i) a $\n^a$ which converts the initial $l=0$
state to a $l=1$ state, and  (ii) an $\eta_i\eta_j$ term  (or its
conjugate), which converts the initial $F_\eta=2$ state to a 
$F_\eta=4$ (or 0, respectively). For such fermion states,
(see {\it definition of} $\alpha$  following \eq{psi-alpha}),
we have  $\alpha=l+2$ which $=3$ in this case.

However, although we do not have a divergent matrix element, we 
have a sum over an infinite number of intermediate states,
given by
\be
- M \sum_{r=0}^\infty \frac{C^2_1(r)}{E_1(r)} = 
- M\sum_{r}^{r_{max}}
\big(A_1 + \frac{B_1}{r} + O(\frac1{r^2})\big) \sim 
- M \big(A_1 r_{max} + B_1 \ln r_{max}
+ {\rm finite}\big)
\label{v1-div}
\ee
since $E_1(r) \equiv E_{int}(r) -E_0
\sim r, C_1(r) \sim \sqrt r$ 
for large $r$.

\subsubsection{Cancellation between $V_1$ and $V_2$ contributions}

The two divergences \eq{v2-div} and \eq{v1-div} are
hard to compare since the cut-off's are different. There is
a way, however, to recast the $V_2$ calculation using the
$r_{max}$ cut-off. By using (see \eq{A6}) $\{Q^{-}_i,
Q^{-j}\}= -P^-\delta^j_{i}$ we can write the $V_2$ term 
in \eq{delta-e} as 
\[
\frac14 {\sum_{int}}' 
| \langle \Psi_0 | \sum_{n} q^{-i}_{n+1,n} | \Psi_{int} 
\rangle|^2
\]
In a manner similar to \eq{v1-div}, the matrix element is now finite,
say $C_2(r)$ (which includes a finite $z$-integral !)
\be
M \sum_{r=0}^\infty {C_2^2(r)} = 
M\sum_{r}^{r_{max}}
\big(A_2 + \frac{B_2}{r} + O(\frac1{r^2})\big) \sim 
M \big(A_2 r_{max} + B_2 \ln r_{max}
+ {\rm finite}\big)
\label{v2-div-new}
\ee
As we will find in Appendix C, $A_1=A_2, B_1=B_2$. 

This implies that the divergent contributions from $V_2$ and
$V_1$ in \eq{delta-e} cancel, leaving a finite result
\footnote{It might seem surprising at first that there 
is a cancellation of divergences between two rather different looking
terms. This is, however, standard in supersymmetric theories. E.g. in
a theory with one scalar superfield with $\Phi^3$ interaction the mass
corrections cancel between a second order perturbation coming from
$V_1=g \bar\psi \psi \phi$ and a first order perturbation coming from
$V_2=g^2 \phi^4$.}. Since for the $V_2$ contribution, we have
calculated the same quantity using the $z_{min}$ cut-off as well as
the $r_{max}$ cut-off, these two cut-offs are clearly related
\[
\frac1{z_{min}^2} \sim r_{max}
\]
reflecting a UV-IR relation. This in particular implies that the
divergence from both $V_1$ and $V_2$ can be interpreted as coming from
strings near $z=0$. 

\subsubsection{Other states}

In the above we have considered as initial state a tensor product of
specific single-bit states. Our choice was prompted by the maximally
divergent $z_{min}$ behaviour in the $V_2$ contribution. For instance,
instead of an $F_\eta=2$ state, if we take $F_\eta=1$ or $F_\eta=3$
states, we will have a $\ln z_{min}$ divergence. In fact these two
cases are the only ones with a potential divergence. None of the
infinite number of other states, with additional $s, \bar s$ or $r$
excitations, or $F_\eta=0 $ or $F_\eta=4 $, have any divergent
$z$-integration.  The methods we have employed work pretty much the
same way for the other divergent initial states. 
Calculations to check the cancellation of divergences in
these remaining cases are in progress and will be
reported elsewhere. 

\subsubsection{Summary of this subsection} 

The fact that the $z\to 0$ divergences cancel implies that our
postulate of treating the string world sheet theory perturbatively
around $\lambda=0$ is the correct one. It is clear that in a purely
bosonic model without fermionic terms such a cancellation would not
have occurred. We believe that the cancellation will persist to all
orders of perturbation theory. It is clearly important to 
explicitly check this.

\section{\label{string-spectrum}String spectrum and AdS/CFT}

In this section we will try to construct string states out of the
multi-bit states discussed above. Our attempts to do so and to compare
with the boundary gauge theory will turn out to be unsuccessful, for
reasons which are discussed below.

The only multi-bit states that are relevant in a straightforward
continuum limit are the ones with an infinite number of bits. All such
states have infinite energy. Consider, for example, the state in
\eq{basic-multi-bit}. For a large number of bits $m$, it is a highly
degenerate state, with the energy diverging as $2m$. A simple ``normal
ordering'' prescription that gets rid of this divergence does not
work. This is because these states satisfy a unitarity bound, which is
saturated by the symmetric state in \eq{sym-wave} which transforms as
$(2m,0,0| \mathbf{0,2m,0})$ under the superconformal group. For
example, subtracting out the divergence from the energy will violate
the unitarity bound for this state.  In fact, it is easy to see that
this is generically the case. One needs a much more sophisticated
normal ordering prescription which depends on the $SO(6)$ quantum
numbers of the state in such a way that the finite part in energy left
after the subtraction continues to satisfy the unitarity bound. This
requirement is equivalent to preserving the superconformal algebra in
the normal ordered theory since the unitarity bounds follow from this
algebra. It is not obvious how to do this. At any rate, we have so far
not succeeded in finding such a normal ordering prescription.

This difficulty in the construction of continuum strings motivates us
to consider the alternative possibility of an essential discreteness
to strings in AdS at zero radius. Thus we may consider the possibility
of {\it defining} string states as the multi-bit states constructed in
the previous section with a {\it finite} number of bits. In this case
strings with any number of bits must be considered since even if to
start with all strings consist of the same number of bits, their
splitting and joining will give rise to other strings with smaller as
well as larger number of bits. Thus we must consider strings made up
of $1, 2, 3, \cdots, \infty$ bits. In the following we will consider such
string states and compare their properties with the corresponding
operators in the boundary gauge theory. Like in the previous section,
the states/operators will be labeled by their energy/dimension ($E_0$
value). As discussed below, it turns out that the spectrum of the
multi-bit states does {\it not} agree with the spectrum of operators
in the boundary gauge theory.

\myitem 1. {\bf Dimension 1}: There are no gauge invariant operators
of dimension 1 (e.g. Tr $F_{\mu\nu}=0$, since we are dealing with
SU(N) gauge theory). On the string theory side too, there is no state
of dimension 1.

\myitem 2. {\bf Dimension 2}: String states of dimension 2
are provided by single bit strings (which correspond to $m=1$ in the
notation of point (2) in section \ref{string-spectrum}). As we have
remarked in section \ref{1bit-comparison}, such states simply form the
supergravity multiplet in AdS$_5\times S^5$ background. The lowest
dimension state of this multiplet \eq{basic-state} corresponds to
dimension 2.  The states in the supergravity multiplet have been
extensively studied \cite{MAGOO} and shown to be in one-to-one
correspondence with the local, gauge-invariant, single-trace chiral
operators of the boundary gauge theory, establishing a complete
equivalence of such Yang-Mills operators with single-bit string states
in our theory. In particular, the state \eq{basic-state} corresponds
to the dimension 2, traceless, symmetric tensor Tr $\phi^a \phi^b$.

\smallskip

\noindent {\bf Missing non-BPS states in the string description:} Besides the
chiral operators, however, there are non-chiral gauge invariant
operators such as Tr $\phi^a\phi^a$ \footnote{In the usual discussion
of AdS supergravity vs CFT at large 'tHooft coupling, one normally
ignores these non-chiral operators because they have large anomalous
dimensions which corresponds to large curvature corrections in
supergravity and hence such states are not considered accessible in
AdS supergravity which is concerned with small
curvature. We are, however, dealing with strings at small
radius AdS or small $\lambda$ for which the anomalous dimensions
of non-chiral operators become small; thus we must
address these states.}, which fall in the SU(2,2$|$4)
multiplet (2,0,0 | 0,0,0). Unfortunately, we do not have any such
normalizable state in the string spectrum \footnote{%
There is a non-normalizable state carrying these quantum numbers.}.

\myitem 3. {\bf Dimension 3}: We found that at dimension 2, the BPS
counting matches between string theory and gauge theory.  How about
dimension 3? We find that the matching continues there: Consider the
gauge theory operator $O_3=A_{(abc)|}{\rm Tr} (\phi^a \phi^b\phi^c)$,
which has the PSU(2,2$|$4) quantum numbers (3,0,0$|${\bf 50}. With
these quantum numbers this is the only gauge invariant operator, since
in a SU(N) YM theory, we cannot for example have
$O_{1,2}=A_{(abc)|}{\rm Tr} (\phi^a){\rm
Tr}(\phi^b\phi^c)$. Interestingly the story is similar in our string
construction. As we remarked above, corresponding to $O_3= $
(3,0,0$|${\bf 50}), there {\em is} a single-bit string state.  It
turns out that this is the only state, since there cannot be a
multi-bit state because even for $m=2$ bits, the minimum value of
$E_0$ will be 4, one higher than that of $O_3$.

\smallskip

\noindent {\bf Non-BPS}: The mismatch vis-a-vis non-BPS states continues at
dimension 3, however. For instance, there is no string state
corresponding to Tr $\phi^a F_{\mu\nu}^a$.

\myitem 4. {\bf Dimension 4 and above--- extra states in string theory}:
At dimension 4, we have the following two chiral gauge theory
operators in gauge theory $O_4=A_{(abcd)|}{\rm Tr} (\phi^a \phi^b\phi^c
\phi^d )$ and $ O_{2,2}=A_{(abcd)|}{\rm Tr} (\phi^a
\phi^b){\rm Tr}(\phi^c \phi^d )$, both carrying
quantum numbers $(E_0, J_1, J_2 |{\mathtt{SU(4)\
repr.}})=(4,0,0|{\mathbf{105}})$. A dimension 4 state
in the string theory can arise in the following ways:
\hfill\break

(a) one 2-bit dimension 4 state (see $m=2$ in
\eq{sym-wave}: this has a dimension 2, $l=0$ excitation at
each bit)\hfill\break  

(b) tensor product of  two 1-bit, $l=0$ string states (this
is a multiple-string state) \hfill\break  

(c) an $l=1$ excited state of a 1-bit string state.
\hfill\break

\noindent Clearly we have extra states in the string theory. Since the
states (a) and (b) appear to be in natural correspondence with the
operators $O_4$ and $ O_{2,2}$ respectively, the state (c) appears to
be a misfit. Indeed, the state (c) is part of a ``Kaluza-Klein tower''
over a single bit.  Ideally one would have liked to have one state per
bit and a string description to arise after combining various bits,
but this is not the situation we have. In our formalism each bit
behaves like a free superparticle in the AdS background with all the
corresponding states available to it.

Generalization to operators of higher dimensions is straightforward.
At dimension 5, we have $O_5= {\rm Tr} \phi^5, O_{2,3}= {\rm Tr}
\phi^2 \ {\rm Tr} \phi^3 $, and there are again three string states,
(a) a 2-bit string state (a dimension 3 excitation in one bit, and a
dimension 2 in the other), (b) a 2-string state, each with one bit,
where the bits are similar to (a), and (c) a single bit state of
dimension 5. Again (c), belonging to the tower over a string state,
appears as extra. 

We conclude that the spectrum of string states considered as multi-bit
states with finite number of bits does not agree with the spectrum of
the operators in the boundary gauge theory.

\section{Discussion and concluding remarks} 

In this paper we have begun the program of quantizing string theory in
AdS at small radius. We make essential use of the light cone gauge. In
this gauge, strings at zero radius seem to fall apart into discrete
bits. We have computed the free spectrum of multi-bit states and
demonstrated that the leading corrections to this spectrum, coming
from a non-zero value of the radius, are finite. Potential divergences
which arise from strings near the boundary $z=0$ cancel because of
bose-fermi cancellations.

Our central aim is to establish a working framework of string theory
in AdS at small radius as a perturbative expansion around zero
radius. Although we have made some progress towards this goal, the
most important problem of obtaining the spectrum of free strings at
zero radius remains unsolved. The importance of a resolution of this
issue cannot be overstated.  One immediate application of these ideas
would then be to test the strong (arbitrary $\lambda$) version of Maldacena
conjecture.  The strong version of the conjecture allows definition of
non\-perturbative string theory (in AdS) in terms of gauge theory.
 
We also have here the beginnings of a calculable framework of strings
coupled to a noncompact space whose gravity description approaches a
singular limit. This example is more non-trivial than that of a string
theory in backgrounds where the compact part of the spacetime includes
an orbifold or a conifold singularity. The singularity here is not
only in a finite codimensional submanifold in space, but the entire
spacetime collapses in the sense that the (constant) curvature blows
up uniformly all over spacetime. Calculation of string theory
correlation functions in such a background should be interesting.

Finally, the success of the light-cone gauge, at least perturbatively,
encourages its use in other backgrounds which have physical gauge
theory duals, especially those which describe a confining gauge theory
like \cite{KleStr}. 

We end by mentioning that it will clearly be important to eventually
extend the present framework to take into account string-string
interactions, which we have ignored in this paper.

\gap5

\noindent{\bf Acknowledgments}

\gap2

\noindent 
One of us (GM) would like to to acknowledge many useful discussions
with Sumit Das, Rajesh Gopakumar, Juan Maldacena, Shiraz Minwalla,
Lubos Motl, Joe Polchinski and Soo-Jong Rey. We would like to thank
Vladimir Dobrev for pointing out an error in an earlier version
of the paper.  Part of this work was done during the String Theory
workshop (December, 2002) at the Harish Chandra Research Institute,
Allahabad. We would like to thank the organizers of the meeting for
providing a stimulating atmosphere and for their excellent
hospitality. One of us (SRW) would like to thank the ASICTP
for hospitality during part of this work.

\gap5

\noindent{\bf Dedication}

\gap2

\noindent This work is dedicated to the memory of Professor Bunji
Sakita, who passed away on 31st August 2002. Prof. Sakita made many
pioneering contributions to high energy physics and in particular to
string theory. His `Reminiscences' \cite{Sakita} are a striking
summary of an exciting period in high energy physics. He was a great
mentor and a deeply humble and generous man. He also contributed
significantly to the high energy physics effort at the Tata Institute
of Fundamental Research in Mumbai. His obituary can be found at
http://www.cerncourier.com/main/article/42/9/20/3.

\newpage

\appendix

\section{Generators of the superconformal algebra in the 
light-cone gauge}

The bosonic part of the $psu(2,2|4)$ superalgebra is the so$(4,2)$
algebra. In terms of the standard hermitian rotation generators,
$S_{AB}$, this algebra is
\bea
[S_{AB},S_{CD}]=-i(\eta_{BC}S_{AB}-\eta_{BD}S_{AC}+\eta_{AD}S_{BC}
-\eta_{AC}S_{BD}),
\label{A1}
\eea where $A,B, \cdots =-1,0,1, \cdots, 4$ and the metric 
$\eta_{AB}$ is diag $[-1,-1,1,1,1,1]$. The $S_{AB}$
acts on AdS$_5$, viewed as the hyperbola
\be
\eta_{AB} Y^A Y^B = -R^2
\label{hyperb}
\ee
Eqn. \eq{A1} can be recast into the standard conformal
algebra in $(3+1)$-dimensions by the redefinitions (see, for example,
\cite{SM})
\bea 
{\tilde D}=S_{-14}, \quad S_{\mu -1}={1 \over
2}({\tilde P}_\mu+ {\tilde K}_\mu), \quad S_{\mu 4} ={1 \over
2}({\tilde P}_\mu-{\tilde K}_\mu), \quad S_{\mu\nu}={\tilde
J}_{\mu\nu},
\label{A2}
\eea
where $\mu, \nu =0,1,2,3$. We will actually be working with a slightly
different set of generators for the conformal algebra. We will use
anti-hermitian generators for the dilations and rotations, while
retaining hermitian generators for translations and special conformal
transformations. That is, we will set 
\bea
{\tilde D}=iD, \quad {\tilde J}_{\mu\nu}=-iJ_{\mu\nu}. 
\label{A2a}
\eea
It will also be convenient to set 
\bea
{\tilde P}_\mu=-\sqrt{2} P_\mu, \quad {\tilde K}_\mu= -\sqrt{2} 
{\tilde K}_\mu.
\label{A2b}
\eea
In terms of these generators the conformal algebra is
\bea
&& [D, P_\mu]=-P_\mu, \quad [D, K_\mu]=K_\mu, \quad [P_\mu, K_\nu]=
\eta_{\mu\nu}D-J_{\mu\nu}, \nn \\
&& [J_{\mu\nu}, P_\gamma]=\eta_{\nu\gamma}P_\mu-\eta_{\mu\gamma}P_\nu,
\quad [J_{\mu\nu}, K_\gamma]=\eta_{\nu\gamma}K_\mu-\eta_{\mu\gamma}K_\nu, 
\nn \\
&& [J_{\mu\nu},J_{\gamma\tau}]=\eta_{\nu\gamma}J_{\mu\tau}
-\eta_{\nu\tau}J_{\mu\gamma}+\eta_{\mu\tau}J_{\nu\gamma}
-\eta_{\mu\gamma}J_{\nu\tau}.
\label{A3}
\eea 
To fix conventions for our light-cone frame, we define the light-cone
variables as follows.
\[
x^\pm \equiv {(x^3 \pm x^0) \over \sqrt{2}}, \quad x={(x^1+ix^2)
\over \sqrt{2}}, \quad \bar x= {(x^1-ix^2)
\over \sqrt{2}}.
\] 
Similarly 
\[
p_\pm \equiv {(p_3
\pm p_0) \over \sqrt{2}}, \quad p^x={(p_1+ip_2) \over \sqrt{2}}, \quad
p^{\bar x}={(p_1-ip_2) \over \sqrt{2}}.
\] 
The corresponding operator statements are in \eq{A7}.
The metric in the light-cone frame is
\[
\eta_{+-}=\eta_{-+}= \eta_{x\bar x}=\eta_{{\bar x} x}=1.
\] 
The algebra in (\ref{A3}) needs to be supplemented by the supersymmetry
generators, $Q^\pm_i$ and $S^\pm_i$ in the light-cone gauge, to get
the full superconformal algebra. The additional commutators are
\bea
&& [D, Q^\pm_i]=-{1 \over 2}Q^\pm_i, \quad [D, S^\pm_i]=
{1 \over 2}S^\pm_i, \nn \\
&& [J^{+-}, Q^\pm_i]=\pm {1 \over 2}Q^\pm_i, \quad 
[J^{+-}, S^\pm_i]=\pm {1 \over 2}S^\pm_i, \quad
[J^{x\bar x}, Q^\pm_i]=\mp {1 \over 2}Q^\pm_i, \quad 
[J^{x\bar x}, S^\pm_i]=\mp {1 \over 2}S^\pm_i, \nn \\
&& [Q^\pm_i, {J^j}_k]=\delta^j_i Q^\pm_k-
{1 \over 4} \delta^j_k Q^\pm_i, \quad
[S^\pm_i, {J^j}_k]=\delta^j_i S^\pm_k-
{1 \over 4} \delta^j_k S^\pm_i, \nn \\
&& [J^{+x}, Q^{-i}]=Q^{+i}, \quad [J^{-\bar x}, Q^{+i}]=-Q^{-i}, \quad
[J^{-x}, S^{+i}]=-S^{-i}, \quad [J^{+\bar x}, S^{-i}]=S^{+i}, \nn \\
&& [S^\mp_i, P^\pm]=Q^\pm_i, \quad  [S^-_i, P^x]=Q^-_i, \quad
[S^+_i, P^{\bar x}]=-Q^+_i, \nn \\
&&  [Q^\mp_i, K^\pm]=-S^\pm_i, \quad  [Q^-_i, K^{\bar x}]=-S^-_i, \quad
[Q^+_i, K^x]=S^+_i.
\label{A4}
\eea
Here ${J^j}_k$ are the generators of su$(4)$ algebra in hermitian basis,
$({J^i}_j)^\dagger={J^j}_i$,
\bea
[{J^i}_j, {J^k}_l]=\delta^k_j {J^i}_l-\delta^i_l {J^k}_j.
\label{A5}
\eea
The anticommutators are
\bea
&& \{Q^{\pm i} ,Q^\pm_j\}=\pm P^\pm \delta^i_j, \quad   
\{Q^{+i},Q^-_j\}= P^x \delta^i_j, \nn \\
&& \{S^{\pm i} ,S^\pm_j\}=\pm K^\pm \delta^i_j, \quad   
\{S^{+i},S^-_j\}= K^{\bar x} \delta^i_j, \nn \\
&& \{Q^{+i} ,S^+_j\}=-J^{+x} \delta^i_j, \quad   
\{Q^{-i},S^-_j\}=-J^{-\bar x} \delta^i_j, \nn \\
&& \{Q^{\pm i},S^{\mp}_j\}={1 \over 2}(J^{+-}+J^{x\bar x} \mp D)\delta^i_j
\mp {J^i}_j.
\label{A6}
\eea
The rest of the (anti)commutation relations are obtained by using the
hermiticity conditions
\bea
&& P^{\pm \dagger}=P^\pm, \quad P^{x\dagger}=P^{\bar x}, \quad 
K^{\pm \dagger}=K^\pm, \quad K^{x\dagger}=K^{\bar x}, \quad 
(Q^{\pm i})^\dagger=Q^\pm_i, \quad (S^{\pm i})^\dagger=S^\pm_i, \nn \\
&& (J^{\pm x})^\dagger=-J^{\pm \bar x}, \quad (J^{+-})^\dagger=-J^{+-},
\quad (J^{x \bar x})^\dagger=J^{x \bar x}, \quad D^\dagger=-D.
\label{A7}
\eea

\subsection{String realization of the superconformal generators}

Expressions for many of the generators have been obtained in \cite{MTT}
\footnote{Expressions for the rest can be obtained from these by using
the (anti)commutation relations given above.}. All the generators may
be written as integrals over the closed string of the corresponding
density, 
\[
G=\int_0^l d\sigma \ {\cal G}(\sigma),
\] 
where $l$ is the
length of the string. Below we reproduce the light-cone gauge
expressions obtained in \cite{MTT} for these densities.  These are
written in terms of the bosonic transverse co-ordinates, $x(\sigma),
{\bar x}(\sigma), z^a(\sigma)$, their momentum conjugates
$p^x(\sigma), p^{\bar x}(\sigma), p^a(\sigma)$, the superpartners
$\theta^i(\sigma), \eta^i(\sigma)$ and their conjugates.
\bea
{\cal P}^- &=& -{1 \over 2 p^+}\biggl(2 p^x p^{\bar x}+(p^z)^2+
{1 \over z^2} \big[ l^i_jl^j_i +
4\eta_i {l^i}_j \eta^j+(\eta_i\eta^i -2)^2-\frac14\big]\biggr)-
{{\rm T}^2 \over 2p^+ z^4}(2x'\x'+(z_a')^2) \nn \\
&& +{{\rm T} \over z^2 p^+} \eta^i\rho^a_{ij}\n^a(\theta'^j-
i{\sqrt{2} \over z} \eta^j x')
+{{\rm T}\over z^2 p^+} \eta_i{\rho^a}^{ij}\n^a
(\theta'_j+i{\sqrt{2} \over z} \eta_j {\bar x}'), \label{A8} \\
{\cal J}^{+x} &=& -ixp^+, \label{A9} \\
{\cal J}^{+-} &=& -ix^-p^+ +2, 
\label{A10} \\
{\cal J}^{x \bar x} &=& i(x p^{\bar x}-{\bar x}p^x)+
{1 \over 2}(\theta^i\theta_i-\eta^i\eta_i), \label{A11} \\
{\cal D} &=& i(x^-p^++x p^{\bar x}+{\bar x}p^x)+
izp^z -{1 \over 2}, \label{A12} \\
{\cal K}^+ &=& {1 \over 2}(z^2+2x{\bar x})p^+, \label{A13} \\
{\cal K}^x &=& {1 \over 2}z^2 p^x-x(x^-p^++x p^{\bar x}+zp^z
+\frac{i}2 +\frac{i}2 (\theta^i\theta_i+\eta^i\eta_i))
+\frac{\theta^i}{\sqrt{p^+}}{\cal S}^+_i, 
\label{A14} \\
{{\cal J}^i}_j &=& {l^i}_j+\theta^i\theta_j+\eta^i\eta_j-
{1 \over 4}\delta^i_j(\theta^k\theta_k+\eta^k\eta_k), \label{A15} \\
{\cal Q}^+_i &=& \sqrt{p^+}\theta_i, \quad {\cal S}^+_i=\sqrt{p^+ \over 2}
z\eta_i+i\sqrt{p^+}x\theta_i, \label{A16} \\
{\cal Q}^-_i &=& {1 \over \sqrt{2p^+}}(\sqrt{2}p^x \theta_i-
i\eta_i p^z- {1 \over z}(-3\eta_i/2 + \eta^j\eta_j\eta_i-2(\eta l)_i))
\nn\\
&&~~~~~~~~~~~~~~~+
{{\rm T} \over \sqrt{2p^+} z^2}\rho^a_{ij}z^a(\theta'^j-
i{\sqrt{2} \over z} \eta^j x').
\label{A17}
\eea
Notation used above: $\rho^a_{ij}$ and $\rho^{aij}$ are SU$(4)$
Clebsh-Gordon coefficients (SO$(6)$ Dirac matrices in the chiral
representation). Some useful identities involving them are
\bea
&& \rho^a_{ij}\rho^{bjk}+\rho^b_{ij}\rho^{ajk}=2\delta^{ab}{\delta^i}_j, 
\nn \\
&& \rho^a_{ij}=-\rho^a_{ji}, \quad \rho^a_{ij}=(\rho^{aji})^*,
\quad \rho^{aij}\rho^a_{km}=2({\delta^i}_m{\delta^j}_k-
{\delta^i}_k{\delta^j}_m)
\nn\\
&& \rho^a_{ij}=\frac12 \epsilon_{ijkl}\ \rho^{a,kl}
\label{A18}
\eea
Also, ${l^i}_j$ is given by
\be
{l^i}_j ={i \over 8}{[\rho^a,\rho^b]^i}_j l^{ab},\quad 
l^{ab}= \n^a {\cal P}^b - \n^b {\cal P}^a,
\label{A19}
\ee
and satisfies the identities
\bea
{l^i}_m{l^m}_j={1\over4}{l^p}_q{l^q}_p{\delta^i}_j+2{l^i}_j, \quad
[{l^i}_j, {l^k}_m]=\delta^k_j {l^i}_m-\delta^i_m {l^k}_j.
\label{A20}
\eea
In the above formulae, $p^z$ is the momentum conjugate to $z$ in the
spherical polar co-ordinates and ${\cal P}^a$ is the corresponding
quantity for $\n^a$. Also, note that $\n^a {\cal P}^a=0$. 

\noindent\underbar{Canonical (anti)commutation relations}:

\gap1

In order to evaluate the above generators
\eq{A10}-\eq{A17} 
on wavefunctions we need the (anti)commutation relations between the
fundamental variables, which are as follows:
\bea
[x^-_0, p^+] && = i,
\nn\\
~[x(\sigma), p^{\bar x}(\sigma')] && =
[{\bar x}(\sigma), p^{x}(\sigma')]=[z(\sigma), p^z(\sigma')]=
i \delta(\sigma - \sigma'),
\nn\\
~[\n^a(\sigma), {\cal P}^b(\sigma')] && = 
i(\delta^{ab}-\n^a \n^b) \delta(\sigma-\sigma'),
\nn\\
~[{\cal P}^a(\sigma), {\cal P}^b(\sigma')] && = 
-i (\n^a {\cal P}^b-\n^b {\cal P}^a) \delta(\sigma-\sigma'), 
\nn\\
\{ \eta^i(\sigma), \eta_j(\sigma')\} && = 
\{ \theta^i(\sigma), \theta_j(\sigma')\} =
\delta^i_j \delta(\sigma - \sigma')
\label{funda-canon}
\eea

\subsection{Useful linear combinations}

In addition to the above generators, some special linear combinations
turn out to be useful (see \eq{threetwo}.  We have already used
$H_\pm,E_0$ which are given by:
\be
H_+ = - P^- + K^+, \, H_- = P^+ - K^- , E_0 = \frac12 (H_+ + H_-)
\label{e0}
\ee
We list below some more. Consider the $so(4)= su(2) \times su(2)$
subalgebra of $so(4,2)$; let us call its generators $\I_i, \K_i$. We
write these in terms of the above generators as
\bea
\I_3 &&= \half J^{x\bar x} + \frac14 (H_+ - H_-),\;
\K_3= \half J^{x\bar x} - \frac14 (H_+ - H_-)
\nn\\
\I_+ &&= \frac12 (J^{+x} + J^{-x} - P^x + K^x),\;
\K_+ = \half(J^{+x} + J^{-x} + P^x - K^x),
\nn\\
\I_- &&=\half( -J^{+\bar x} - J^{-\bar x} - P^{\bar x} + K^{\bar x}),\;
\K_- =\half( -J^{+\bar x} - J^{-\bar x} + P^{\bar x} - K^{\bar x})
\label{so4}
\eea
Using the above we get
\be
J^{x\bar x}= \K_3+ \I_3,\, H_\pm = E_0 \mp (\K_3 - \I_3)
\ee 
On a state annihilated by $\I_-, \K_+$, we have $\K_3=J_1,
\I_3= -J_2$, therefore
\be
H_\pm = E_0 \mp (J_1 + J_2), \, J^{x\bar x}= J_1 - J_2
\label{j1-j2}
\ee
We also list here some special linear combinations of
the odd generators which are particularly useful:
\bea
&& \widetilde Q^\pm_i= Q^\pm_i \mp S_i^\mp, \;
\widetilde Q^{\pm i}= Q^{\pm i} \pm S^{\mp i}, 
\label{raising}\\
&& \widetilde S^\pm_i= Q^\pm_i \pm S_i^\mp, \;
\widetilde S^{\pm i}= Q^{\pm i} \mp S^{\mp i}
\label{lowering}
\eea
The first line raises the value of $E_0$, while the second lowers the
value of $E_0$, by $1/2$. The bottom of a supermultiplet is defined by
annihilation of the $\widetilde S$ operators ({\em cf.} these are
similar to the $Q',S'$ operators discussed in \cite{SM}.)

\section{Discretized expressions for the generators}

In this appendix we list the discretized versions of the expressions
for the generators given above. As discussed in the main text, section
3, the lattice spacing $\epsilon$ completely disappears from these
expressions after appropriately rescaling the momenta and the
fermionic variables to have finite canonical commutation relations
among the bit variables. These rescalings are 
\bea
(p_n^+, p_n^x,
p_n^{\bar x}, p^z_n, {\cal P}^a_n)
&& = \epsilon  \biggl( p^+(\sigma), p^x(\sigma),
p^{\bar x}(\sigma), p^z(\sigma), {\cal P}^a(\sigma) \biggr), \quad \sigma= n \epsilon
\nn\\
(x^-_0, x_n,
{\bar x}_n, z_n, \n^a_n)
&& =   \biggl( x^-(\sigma), x(\sigma),
{\bar x}(\sigma), z(\sigma), \n^a(\sigma) \biggr), 
\nn\\
(\eta^i_n, \eta_{j,n}, \theta^i_n, \theta_{j,n})
&& =
{\epsilon}^{1/2}\biggl(
\eta^i(\sigma), \eta_{j}(\sigma), 
\theta^i(\sigma), \theta_{j}(\sigma)\biggr)
\label{rescalings}
\eea
The discrete version of the
canonical (anti)commutation relations \eq{funda-canon} is
\bea
&& [x_n, \ p_m^{\bar x}]=i\delta_{nm}, \quad [{\bar x_n},
p_m^x]=i\delta_{nm}, \quad [z_n, p_m^z]=i\delta_{nm}, \nn \\
&& [\n^a_m, {\cal P}^b_n]=i (\delta^{ab}-\n^a_m \n^b_m) 
\delta_{nm}, \nn\\
&& [{\cal P}^a_m, {\cal P}^b_n]= 
-i (\n^a_m {\cal P}^b_m-\n^b_m {\cal P}^a_m) \delta_{nm}, \nn\\
&& \{\eta_n^i, \eta_{jm}\}=\delta_{nm}\delta^i_j, \quad 
\{\theta_n^i, \theta_{jm}\}=\delta_{nm}\delta^i_j.
\label{B1}
\eea
The oscillators for the two transverse directions $x, \bar x$, parallel
to the boundary, are defined as follows:
\bea 
&& a={p^x-ip^+ x \over
\sqrt{2p^+}}, \quad a^\dagger={p^{\bar x}+ip^+ {\bar x} \over
\sqrt{2p^+}}, \quad [a, a^\dagger]=1, \nn \\ && {\bar a}={p^{\bar
x}-ip^+ {\bar x} \over \sqrt{2p^+}}, \quad {\bar a}^\dagger={p^x+ip^+
x \over \sqrt{2p^+}}, \quad [{\bar a}, {\bar a}^\dagger]=1.
\label{B1a}
\eea 

The general form of the discrete version of the generators, in the
limit T $\rightarrow 0$ indicated by the subscript $'0'$ on the
generators, is $G_0=\sum_{n=1}^M g_{0n}$, where $g_{0n}$ is the
corresponding generator for the $n$th bit. We list the latter below,
but omit the bit index for ease of notation.
\bea
 p^-_0 &=& -{1 \over 2 p^+} \biggl(2p^xp^{\bar x}+
(p^z)^2 +{1 \over z^2}((l^i_j l^j_i)+4\eta_{i} {l^i}_j \eta^j+
(\eta_i\eta^i-2)^2-\frac14)\biggr), \label{B2} \\
j_0^{+x} &=& -ixp^+, \label{B3} \\
j_0^{+-} &=& -ix^-p^+ +2, 
\label{B4} \\
j_0^{x \bar x} &=& i(x p^{\bar x}-{\bar x}p^x)+
{1 \over 2}(\eta_i\eta^i-\theta_i\theta^i), \label{B5} \\
d_0 &=& i(x^-p^++x p^{\bar x}+{\bar x}p^x)+
izp^z-{1 \over 2}, \label{B6} \\
k_0^+ &=& {1 \over 2}(z^2+2x{\bar x})p^+, \label{B7} \\
k_0^x &=& {1 \over 2}z^2 p^x-x\biggl(x^-p^++x p^{\bar x}+zp^z +\frac{i}2
+\frac{i}2(\theta^i \theta_i +\eta^i \eta_i)\biggr)
+ \frac{1}{\sqrt{p^+}} \theta^i s^+_{0i}, 
\nn\\
\label{B8} \\
{j_0^i}_j &=& {l^i}_j+\theta^i\theta_j+\eta^i\eta_j-
{1 \over 4}\delta^i_j(\theta^k\theta_k+\eta^k\eta_k), \label{B9} \\
q^+_{0i} &=& \sqrt{p^+}\theta_i, \quad s^{+}_{0i}=\sqrt{p^+ \over 2}
z\eta_i+i\sqrt{p^+}x\theta_i, \label{B10} \\
q^-_{0i} &=& {1 \over \sqrt{2p^+}}(\sqrt{2}p^x \theta_i -
i\eta_i p^z- {1 \over z}(-3\eta_i/2 +
\eta^j\eta_j\eta_i-2(\eta l)_i)). \label{B11}
\eea  
Others can be obtained using the (anti)commutation relations.

When T is non-zero some of the generators \footnote{These are $P^-,
K^-, J^{-x}, J^{-\x}, Q^{-i}$ and $S^{-i}$, the so-called
dynamical generators. Note that they also receive corrections from
string-string interactions, which we are ignoring here.} get modified by
terms that involve bit-bit interactions. We write these generators as
$G=G_0+G_1$, where $G_1=\sum_n g_{1(n+1,n)}$ vanishes for T$=0$. We
list $g_{1(n+1,n)}$ terms for some of the generators that get
modified.  
\bea
-p^-_{1(n+1,n)} &=& {{\rm T}^2 \over 2p^+}\biggl(2
{|x_{n+1}-x_n|^2 \over z_n^4}+{(z^a_{n+1}-z^a_n)^2 
\over z_{n+1}^2z_n^2}\biggr) 
- {{\rm T} \over z_n^2 p^+}\eta^i_n\rho^a_{ij}\n_n^a 
\biggl((\theta_{n+1}^j-\theta_n^j)-\biggr. \nn\\
&& i{\sqrt{2} \over z_n} \eta_n^j (x_{n+1}-x_n) \biggl.\biggr) 
+{{\rm T}\over z_n^2 p^+} \eta_{ni}{\rho^a}^{ij}\n_n^a
\biggl((\theta_{n+1j}-\theta_{nj})+i{\sqrt{2} \over z_n} 
\eta_{nj} (\x_{n+1}-\x_n) \biggr), 
\nn\\
\label{B12} \\
q^-_{1i(n+1,n)} &=& {{\rm T} \over \sqrt{2p^+}}\rho^a_{ij} 
\biggl({\n^a_n \over z_n} (\theta^j_{n+1}-\theta^j_n)
-i\sqrt{2}{\n^a_n \over z^2_n}\eta^j_n(x_{n+1}-x_n)\biggr).
\label{B13}
\eea
These expressions are obtained by a straightforward discretization of
the corresponding continuum expressions in (\ref{A8}) and
(\ref{A17}). There is of course no a$'$ priori unique discretized
expression. We make a natural choice for a minimal set and then fix
the rest by demanding that the algebra be satisfied. Thus, e.g.,
\eq{B12} can be independently obtained from \eq{B13}
using \eq{A6}: $\{ Q^{-i}, Q^-_j \} = -P^- \delta^i_j$.

\section{Perturbation of $H_+$ spectrum and
cancellation of  $z\to 0$ divergence}

We assume the string to consist of $M$ bits. 
We define  $H_+ = H^0_+ + T V_1 + T^2 V_2$:
\bea
H^0_+ 
= \sum_{n=1}^M
\biggl[
-\frac{\del_{x_n} \del_{\bar x_n}}{p^+} + p^+ x_n \bar x_n 
- && {\del_{z_n}^2 \over 2p^+}+{p^+ \over 2}z^2_n 
+{1 \over 2p^+z^2_n} 
\big(
({l^i}_{n,j})^2+4\eta_{n,i} 
{l^i}_{n,j} \eta^j_n+
(\eta_{n,i}\eta^i_n-2)^2-{1 \over 4} 
\big) 
\biggr]
\nn\\
V_2 
= \sum_{n=1}^M
\biggl[
{1 \over 2p^+}
\biggl(
2{|x_{n+1}-x_n|^2 \over z_n^4}+ && {(z^a_{n+1}-z^a_n)^2 
\over z_{n+1}^2z_n^2}
\biggr)  
\biggr]
\nn\\
V_1 =\sum_{n=1}^M v_{1(n+1,n)}
\equiv \sum_{n=1}^M
\biggl[&&\biggr. - 
{{1} \over z_n^2 p^+}\eta^i_n\rho^a_{ij}\n_n^a 
\biggl(\biggr. 
(\theta_{n+1}^j-\theta_n^j) -
i{\sqrt{2} \over z_n} \eta_n^j (x_{n+1}-x_n) 
\biggl.\biggr) \nn\\
&& +{{1}\over z_n^2 p^+} \eta_{ni}{\rho^a}^{ij}\n_n^a
\biggl(
(\theta_{n+1j}-\theta_{nj})+i{\sqrt{2} \over z_n} 
\eta_{nj} (\x_{n+1}-\x_n) 
\biggr)
\biggl.\biggr]
\label{h0-v1-v2}
\eea 
\subsection{Calculation of $\langle \Psi_0 | V_2 | \Psi_0 
\rangle$}

We will calculate this quantity as
\bea
\langle \Psi_0 | V_2 | \Psi_0 \rangle 
= &&
\frac14  \langle \Psi_0 |\{ Q_{1,i}^-, Q^{-i}_1 \} | \Psi_0 \rangle 
\nn\\
= \frac14 \sum_{n,m}\sum_{int} && \biggl[\biggr.
\langle \Psi_0 | q^-_{(n+1,n)i} | \Psi_{int} \rangle 
\langle \Psi_{int} | q^{-i}_{(n+1,n)} | \Psi_0 \rangle
\nn\\ 
&&  +\langle \Psi_0 | q^{-i}_{(n+1,n)} | \Psi_{int} \rangle 
\langle \Psi_{int} | q^-_{(n+1,n)i} | \Psi_0 \rangle \biggl.\biggr]
\label{v2-vev}
\eea
The first equality can be derived by considering the $T^2$ term in the
equality $-P^- = \frac14 \{ Q_{i}^-, Q^{-i} \}$.
Here, 
\be      
q^-_{(n+1,n)i}= \frac1{\sqrt 2 z_n} (\rho.\n_n)_{ij}
\left[ (\theta^j_{n+1} - \theta^j_n) + 
\frac1{z_n} \eta^j_n \left(  (a_{n+1} - {\bar a}^\dagger_{n+1}) -
(a_{n} - {\bar a}^\dagger_{n}) \right) \right]
\label{q-}
\ee
This expression is as in \eq{B13} except that we have
rescaled $\sqrt{p^+} z_n \to z_n, \sqrt{p^+} x_n \to x_n$
and have used $-i \sqrt 2 x_n = a_n - {\bar a}^\dagger_{n}$.

Since the intermediate states $ | \Psi_{int} \rangle $ involved for
various terms in \eq{q-} are different, we can separately treat the
contribution of each term in \eq{q-} to \eq{v2-vev}. Let us
consider, e.g., the term
\be \frac1{\sqrt 2 z_n} (\rho.\n_n)_{ij} \theta^j_{n+1}
\label{sample}
\ee
The intermediate state that will click here is represented in the
figure below. 
\noindent
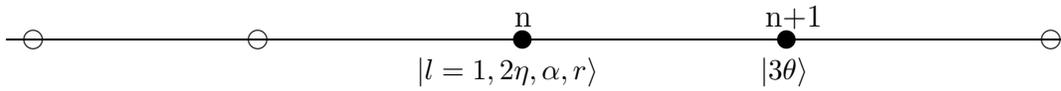
\begin{picture}(500,70)(1,1) 
\put(1,50){ \line(1,0){400}}
\put(15,50){\circle{7}}
\put(100,50){\circle{7}}
\put(197,55){n}
\put(200,50){\circle*{7}}
\put(160,35){\small $\ket{l=1,2\eta,\alpha,r}$}
\put(292,55){n+1}
\put(300,50){\circle*{7}}
\put(290,35){\small $\ket{3\theta}$}
\put(400,50){\circle{7}}
\end{picture}
\noindent
The wavefunction in the picture differs from the initial state only in
the $n$-th and the $n+1$-th bits (the solid circles) and only
in the quantum numbers exhibited \footnote{The full
$\ket{\Psi_{int}}$ is a cyclic permutation of such wavefunctions.}
($\ket{2\eta}$ denotes $\ket{F_\eta=2}$,
etc). Explanation for the $n$-th bit: $l=1$ appears because $\n^a_n$
carries $l=1$, the values of $\alpha$ can  be $l, l+2$ or $l+4$,
i.e. 1,3 or 5, depending on whether $\ket{2\eta}$ belongs to $P_2
\V_2, Q_2\V_2$ or $R_2 \V_2$ (see \eq{p2} and below).
Only the quantum number $r$ remains
unspecified, any $r=0,1,2...$ clicks.  The corresponding matrix
element of the $z$-wavefunctions is (cf. $C_2(r)$ in \eq{v2-div-new})
\be
C_2^{2\eta,\alpha}(r)=
\int_0^\infty  dz \psi^0_0(z)  \frac1{\sqrt 2 z} \psi^\alpha_r(z)
= \frac1{\sqrt 2} \frac{(r + {\alpha-1 \over 2})!}{r!}
\sqrt{\frac{r!}{(r+\alpha)!}}
\label{c-alpha-r}
\ee
Having done the $z$-(and the trivial $x, \bar x$-)integration let us
now leave the S$^5$ and the fermion matrix elements unevaluated. This
amounts to finding an {\em effective interaction} in the finite
dimensional S$^5$ and fermionic sector, after integrating out the
bosonic degrees of freedom $z, x, \x$.  After carrying out the sum
$\sum_r$ involved in the Eqn. \eq{v2-vev} for each $\alpha=1,3$ or 5,
we get the following contribution of the interaction \eq{sample} to
\eq{v2-vev}:
\be
\frac14 \sum_{n,m}\sum_{\alpha=1,3,5} C_2^{2\eta,\alpha}
\langle \tilde \Psi_0 |  (\rho.\n_n)_{ij} \theta^j_{n+1}|
\tilde \Psi_{int,(2\eta,\alpha)} \rangle
\langle \tilde \Psi_{int,(2\eta,\alpha)} 
|  (\rho.\n_m)^{ki} \theta_{m+1,k}|
\tilde \Psi_{0} \rangle
\label{effective}
\ee
where
\be
C_2^{2\eta,\alpha} = \sum_{r=0}^\infty \left(C_2^{2\eta,\alpha}(r)
\right)^2
\label{c-2eta-alpha}
\ee
In the above, the tildes on the wavefunctions denote that they
depend now only on the $S^5$ and the fermionic coordinates.  The
$(2\eta,\alpha)$ tag on the $\tilde \Psi_{int}$ reminds us that 
we must choose the
appropriate projection ($P_2, Q_2$ or $R_2$) on the wavefunction
$\ket{l=1,2\eta}$ on whichever bit it appears.

What remains now is to compute the equivalent of \eq{effective}
for all the other interaction terms. We will write here only one
other term in \eq{q-} in some detail, viz.
\be
\frac1{\sqrt 2 z_n} (\rho.\n_n)_{ij} \eta^j_{n} a_{n+1}
\label{sample-2}
\ee
The pictorial representation of the intermediate state now is

\noindent
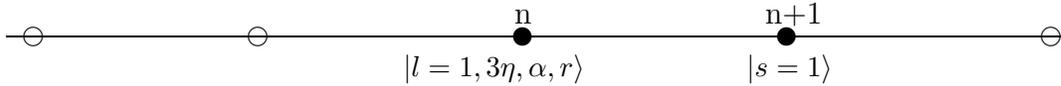
\begin{picture}(500,70)(1,1) 
\put(1,50){ \line(1,0){400}}
\put(15,50){\circle{7}}
\put(100,50){\circle{7}}
\put(197,55){n}
\put(200,50){\circle*{7}}
\put(155,35){\small $\ket{l=1,3\eta,\alpha,r}$}
\put(292,55){n+1}
\put(300,50){\circle*{7}}
\put(285,35){\small $\ket{s=1}$}
\put(400,50){\circle{7}}
\end{picture}

\noindent
Once again only the changed quantum numbers are shown. $s=1$ means a
harmonic oscillator excitation of the '$a$' type (see \eq{psi-alpha}).
The $\alpha$ value of $\ket{l=1,3\eta}$ is determined ($\alpha=l+1=2$
or $\alpha=l+3=4$), depending on whether the $F_\eta=3$ 
state belongs to $P_3
\V_3$ or $Q_3 \V_3$.
The $z,x,\bar x$ integration gives
\[
C_2^{3\eta,\alpha}(r)=
\int_0^\alpha dz \psi^0_0(z)  \frac1{\sqrt 2 z} \psi^\alpha_r(z)
= \frac{\sqrt 2}{\alpha} 
\frac{(r + {\alpha \over 2})!}{\sqrt{r!(r+\alpha)!}}
\]
The contribution to \eq{v2-vev} in terms of $S^5$ and fermions
is, in this case,
\be
\frac14 \sum_{n,m}\sum_{\alpha=2,4}C_2^{3\eta,\alpha}
\langle \tilde \Psi_0 |  (\rho.\n_n)_{ij} \eta^j_{n} a_{n+1}|
\tilde \Psi_{int,(3\eta,\alpha)} \rangle
\langle \tilde \Psi_{int,(3\eta,\alpha)} |  
(\rho.\n_m)^{ki} \eta_{m+1,k}|
\tilde \Psi_{0} \rangle
\label{effective-2}
\ee
where
\be
C_2^{3\eta,\alpha}= \sum_{r=0}^\infty \left(C_2^{3\eta,\alpha}
\right)^2
\label{c-3eta-alpha}
\ee
From these examples, it is clear how to compute 
equations like \eq{effective-2} and \eq{c-3eta-alpha} for all
the other terms in \eq{v2-vev}.

\myitem \underbar{Divergence}:

If we explicitly evaluate $C_2^{2\eta,\alpha},C_2^{3\eta,\alpha}$ using
their definitions, we find that
\bea
C_2^{2\eta,\alpha}
&& = \half \sum_{r=1}^\infty \frac1{r} + F_2(\alpha)
\nn\\
C_2^{3\eta,\alpha} 
&& = A'_2(\alpha) \sum_{r=1}^\infty 1
+ B'_2(\alpha) \sum_{r=1}^\infty \frac1{r} + F'_2(\alpha)
\eea
Thus the two types of divergences that appear here are
$\sum_{r=1}^{r_{max}} 1$ and $\sum_{r=1}^{r_{max}} \frac1{r}$, where
we have replaced the upper limit $r=\infty$ by a cut-off $r_{max}$.
The coefficients of such divergences in \eq{v2-vev} are calculated by
combining equations like \eq{effective} and \eq{effective-2}.

\subsection{The $V_1$ contribution}

This term has an expression
\be
- \frac14 \sum_{n,m}\sum_{int}\frac1{E_{int}-E_0} \biggl[
\langle \Psi_0 | v_{1(n+1,n)} | \Psi_{int} \rangle 
\langle \Psi_{int} | v_{1(n+1,n)} | \Psi_0 \rangle\biggr]
\label{v1-term}
\ee
which is similar to \eq{v2-vev}, except that it has
an additional factor due to the
energy denominator.

Like in the case of the $V_2$ term in the previous subsection,
we will  consider each term of $v_{1(n+1,n)}$ in turn
(see \eq{h0-v1-v2}), consider the intermediate
states that click in each case, and ``integrate out''
the $z, x, \x$ variables, leaving matrix elements depending
on $S^5$ and fermions.

As an example we consider
\[
{1 \over z_n^2 p^+}\eta^i_n(\rho.\n)_{ij} \theta_{n+1}^j
\]
The intermediate state is represented by the diagram

\noindent
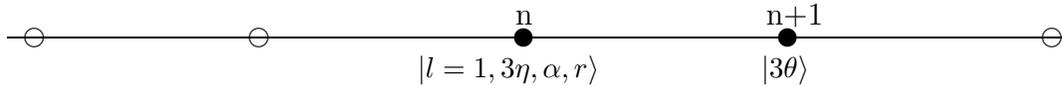
\begin{picture}(500,70)(1,1) 
\put(1,50){ \line(1,0){400}}
\put(15,50){\circle{7}}
\put(100,50){\circle{7}}
\put(197,55){n}
\put(200,50){\circle*{7}}
\put(160,35){\small $\ket{l=1,3\eta,\alpha,r}$}
\put(292,55){n+1}
\put(300,50){\circle*{7}}
\put(290,35){\small $\ket{3\theta}$}
\put(400,50){\circle{7}}
\end{picture}

\noindent
The values of $\alpha$ here are $l+1=2$ or $l+3=4$.
The $z$-integral involved in the matrix element
is $C_1^{3\eta,\alpha}(r)$, which 
is identical to $C_2^{3\eta,\alpha}(r)$. After performing the
$r$-sum we get the following contribution
\be
\frac14 \sum_{n,m}\sum_{\alpha=2,4}C_1^{3\eta,\alpha}
\langle \tilde \Psi_0 | \eta^i_n (\rho.\n_n)_{ij} \theta^j_{n+1}|
\tilde \Psi_{int,(3\eta,\alpha)} \rangle
\langle \tilde \Psi_{int,(3\eta,\alpha)} |  
(\rho.\n_m)^{kl} \theta_{m+1,k}\eta_{n,l}|
\tilde \Psi_{0} \rangle
\label{effective-3}
\ee
where
\be
C_1^{3\eta,\alpha}= \sum_r \frac{\left[ C_2^{3\eta,\alpha}(r)\right]^2}{
(2r + \alpha + 2)-2}
\label{c-3eta-alpha-1}
\ee
The divergence of this term is of the type $\sum_r^{r_{max}} 1/r$.

\subsection{Cancellation of Divergence}

We now collect the coefficient of $\sum_r^{r_{max}} 1$ and
$\sum_r^{r_{max}} (1/r)$ of all the terms in \eq{v2-vev}
and \eq{v1-term}. We will explicitly write here only the
coefficient of $\sum_r 1$. This is given by
\bea
&& \frac14\sum_{n,m}
\biggl[\biggr.
\langle \tilde \Psi_0 |  (\rho.\n_n)_{ij} \eta^j_{n} |
\tilde \Psi_{int,(3\eta,\alpha=2)} \rangle
\langle \tilde \Psi_{int,(3\eta,\alpha=2)} |  
(\rho.\n_m)^{ki} \eta_{m,k}|
\tilde \Psi_{0} \rangle
\nn\\
&&~~~~~~~~~~+  \frac14 
\langle \tilde \Psi_0 |  (\rho.\n_n)_{ij} \eta^j_{n} |
\tilde \Psi_{int,(3\eta,\alpha=4)} \rangle
\langle \tilde \Psi_{int,(3\eta,\alpha=4)} |  
(\rho.\n_m)^{ki} \eta_{m,k}|
\tilde \Psi_{0} \rangle \biggl.\biggr]
\nn\\
&&\frac14\sum_{n,m}
\biggl[\biggr.
\langle \tilde \Psi_0 |  (\rho.\n_n)^{ij} \eta_{n,i} |
\tilde \Psi_{int,(1\eta,\alpha=2)} \rangle
\langle \tilde \Psi_{int,(1\eta,\alpha=2)} |  
(\rho.\n_m)_{jk} \eta_m^{k}|
\tilde \Psi_{0} \rangle
\nn\\
&&~~~~~~~~~~+ \frac14 
\langle \tilde \Psi_0 |  (\rho.\n_n)^{ij} \eta_{n,i} |
\tilde \Psi_{int,(1\eta,\alpha=4)} \rangle
\langle \tilde \Psi_{int,(1\eta,\alpha=4)} |  
(\rho.\n_m)_{jk} \eta_m^{k}|
\tilde \Psi_{0} \rangle \biggl.\biggr]
\nn\\
 && \!\!\! - \frac14\sum_{n,m}
\langle \tilde \Psi_0 | \eta_n^i (\rho.\n_n)_{ij} \eta_{n}^i |
\tilde \Psi_{int,(4\eta,\alpha=3)} \rangle
\langle \tilde \Psi_{int,(4\eta,\alpha=3)} |  
\eta_{m,k}(\rho.\n_m)^{kl} \eta_{m,l}|
\tilde \Psi_{0} \rangle 
\nn\\
&&  \!\!\!  - \frac14\sum_{n,m}
\langle \tilde \Psi_0 | \eta_{n,i} (\rho.\n_n)^{ij} \eta_{n,i} |
\tilde \Psi_{int,(0\eta,\alpha=3)} \rangle
\langle \tilde \Psi_{int,(0\eta,\alpha=3)} |  
\eta_{m,k}(\rho.\n_m)_{kl} \eta_{m,l}|
\tilde \Psi_{0} \rangle
\label{coeff-rmax}
\nn\\
\eea
By repeatedly using (a) the fundamental anticommutation relations of
the fermions \eq{B1}, (b) the explicit form of the projection operators
$P_1, Q_1, P_3$ and $Q_3$ (see \eq{p1}, \eq{p3}), and (c)
$\n^a$-space vev's such as
\[
\langle \n^a \n^b \rangle = \frac16 \delta^{ab},
\quad \langle \n^a l^i_j \n^b \rangle=\frac1{24}[ \rho^a, \rho^b]^i_j
\]
we find that  the above term \eq{coeff-rmax} indeed
vanishes. Note that the first two terms (with
positive sign) arise from $V_2$ the last two (negative
sign) from $V_1$.

The coefficient of $\sum_r^{r_{max}} 1/r$ vanishes by a similar, but
somewhat lengthier computation.

\section{Nonlocal features of stringy representation
in the light cone gauge}

Although many generators of the superconformal algebra involve only
the zero mode $x^-_0$ of $x^-(\sigma)$, some of the generators, like
$K^x$, also involve non-zero modes of $x^-(\sigma)$. These generators
have expressions which are non-local in $\sigma$. In general,
therefore, even at $T=0$, a representation of the stringy realization
of the superconformal algebra may be achieved only by taking linear
combinations (over and above the cyclic sum in \eq{fourone}) of naive
tensor product of single-bit states.  In the following we will
consider the simplest possible example of such a state \eq{sym-wave}
where a linear combination ({\it a la} Clebsch-Gordon) is already in
place because we have taken a symmetric and traceless tensor product
of the $SU(4)$ polarizations.

In order to see if such states carry a representation of the
superconformal algebra, including the non-local generators, let us
consider the example of the $so(4)$ subalgebra of the AdS$_5$ group
$so(4,2)$.  As shown in \eq{so4}, the generators of this algebra
involve $K^x$ which contains $\tilde x^-$ and the action of $so(4)$
therefore constitutes a non-trivial example. 

The state \eq{sym-wave} is expected to be a singlet of $so(4)$.
Therefore it must be annihilated by $\K_\pm, \I_\pm$. Since these
generators depend on $x^-$, we need to first construct an appropriate
action of this operator on the states.  Recall that in the light cone gauge 
the classical Gauss law constraint of $\sigma$-reparametrizations is \cite{MTT}
\bea
x^-(\sigma) 
=x(0) - && \int_0^{\sigma} {d\tilde \sigma 
\over2 p^+} 
\biggl( x' p^{\bar x} + \bar x' p^x + z' p^z + {\n^a}' {\cal P}^a
\nn\\
&& + i\big( \eta^i\eta_i' - {\eta^i}'\eta_i +
\theta^i\theta_i' - {\theta^i}'\theta_i  \big)\biggr)(\tilde \sigma).
\label{x-}
\eea
Here $x(0)$ is related to the zero mode $x^-_0 \equiv \int_0^1 
d\sigma~x^-(\sigma)$ by
\bea
x(0)=x^-_0+  \int_0^1 d\sigma \int_0^{\sigma} {d\tilde \sigma 
\over2 p^+} 
\biggl(&& x' p^{\bar x} + \bar x' p^x + z' p^z + {\n^a}' {\cal P}^a
\nn\\
&& + i\big( \eta^i\eta_i' - {\eta^i}'\eta_i +
\theta^i\theta_i' - {\theta^i}'\theta_i  \big)\biggr)(\tilde \sigma).
\label{xzero}
\eea
To discretize a
combination of the form $p^i(\tilde\sigma) x_i'(\tilde \sigma)$, we
replace the local term $p^i(\tilde \sigma)$ by $(p^i_{k+1} +
p^i_k)/2$, the $\tilde \sigma$-derivative by a discrete difference,
i.e., $x'_i(\tilde \sigma) \to \epsilon^{-1}(x_{i,k+1}- x_{i,k})$ and
perform the rescalings \eq{rescalings}.  This leads to

\bea
x^-_{n+1} = x^-_0 &&
-\frac1{2p^+}(1-{1 \over M} \sum_{n=1}^M) \sum_{k=1}^n \biggl[\biggr. 
(x_{k+1}-x_k)(p^{\bar x}_{k+1}+p^{\bar x}_k)+(\bar x_{k+1}-\bar x_k) 
(p^x_{k+1}+p^x_k) \nn\\
&& +(z_{k+1}-z_{k})(p^z_{k+1}+p^z_k)+
(\n^a_{k+1} {\cal P}^a_k-\n^a_k {\cal P}^a_{k+1})
+ i\beta({z_{k+1} \over z_k}-{z_k \over z_{k+1}})
\nn\\
&& + i \biggl\{( \eta^i_{k+1} \eta_{i,k} - 
\eta^i_{k} \eta_{i,k+1})
+ ( \theta^i_{k+1} \theta_{i,k} - 
\theta^i_{k} \theta_{i,k+1}) \biggr\} \biggl.\biggr]
\label{discrete-x-}
\eea
In this expression $\beta$ is a constant. This extra term is expected
in the quantum theory for the action of $x^-(\sigma)$ on
wave-functions $\psi$ in the polar coordinates $z, \n^a$.

To evaluate the action of the $so(4)$ generators $\K_\pm, \I_\pm$ on
the state \eq{sym-wave} we also need a discretized expression for
the generator $J^{-x}$. This can be guessed from the clasical
continuum expression for it, which may be obtained from the known
generators by using the algebra and the basic Poisson brackets. In
this way we get
\bea
J^{-x}=\sum_{n=1}^M \biggl[\biggr. ip^x_n x^-_n - && i x_n p^-_n + 
{1 \over \sqrt{p^+}}\theta^i_nQ^-_{i,n} 
\nn\\ 
&& -\biggl\{1+{1 \over M}+{1 \over 2}\theta^i_n\theta_{i,n}+
{1 \over 2}\eta^i_n\eta_{i,n}\biggr\}{p^x_n \over p^+}\biggl.\biggr]
\label{j-x}
\eea
Using this one can show that, for an appropriate value of the constant
$\beta$ in \eq{discrete-x-}, the $so(4)$ generators $\K_\pm, \I_\pm$
annihilate the state \eq{sym-wave}. Note that in the fermionc part of
the wavefunction, it is important to take the traceless symmetric
linear combination of the fermion polarizations. For instance, one
could consider a state which has only
$\eta_{1,n}\eta_{2,n}\theta_{1,n}\theta_{2,n}$ at each bit $n$.

One needs to check that the value of $\beta$ for which the above is
true is compatible with the superconformal algebra. In fact, it is an
open question whether the discretized algebra is satisfied. It is
likely that the specific prescription for discretization plays an
important role in this. We do not have a complete understanding of
this issue yet and hope to come back to it \cite{DMWp}.

\myitem\underbar{Summary:}

With the above action of $x^-(\sigma)$ on $\ket{\Psi_{\rm sym}}$, it is
easy to see that both $H_-$ and the lowering operators $\widetilde S$
(see \eq{lowering}) annihilate $\ket{\Psi_{\rm sym}}$.


\begin{thebibliography}{999}

\bibitem{Brezin:eb}
E.~.~Brezin and S.~R.~Wadia,
``The Large N Expansion In Quantum Field Theory And
Statistical Physics: From Spin Systems To
Two-Dimensional Gravity,'' World Scientific, 1993.
\bibitem{Polyakov:ca}
A.~M.~Polyakov,
``Gauge Fields As Rings Of Glue,''
Nucl.\ Phys.\ B {\bf 164}, 171 (1980).
\bibitem{Polyakov:1998ju}
A.~M.~Polyakov,
``The wall of the cave,''
Int.\ J.\ Mod.\ Phys.\ A {\bf 14}, 645 (1999)
[arXiv:hep-th/9809057].
\bibitem{Current}
E. Witten, ``Black holes and quark confinement,''
Current Science, Vol  81  No 12, 25th Dec 2001.
http://tejas.serc.iisc.ernet.in/~currsci/dec252001/contents.htm
\bibitem{Mal97} J. Maldacena, ``The large $N$ limit of superconformal
field theories and supergravity'', Adv. Theor. Math. Phys. {\bf 2},
231 (1998) [Int. J. Theor. Phys. {\bf 38}, 1113 (1998)],
hep-th/9711200.
\bibitem{DMW}
J.~R.~David, G.~Mandal and S.~R.~Wadia,
``Microscopic formulation of black holes in string theory,''
Phys.\ Rept.\  {\bf 369}, 549 (2002)
[arXiv:hep-th/0203048].
\bibitem{GKP} S.S. Gubser, I.R. Klebanov and A.M. Polyakov,
``Gauge-theory correlators from non-critical string theory'', Phys. Lett.
{\bf B428}, 105 (1998), hep-th/9802109.
\bibitem{W} E. Witten, ``Anti-de Sitter space and holography'',
Adv. Theor. Math. Phys.{\bf 2}, 253 (1998), hep-th/9802150.
\bibitem{MAGOO} O. Aharony, S.S. Gubser, J. Maldacena, H. Ooguri and
Y. Oz, ``Large $N$ field theories, string theory and gravity'',
Phys. Rept. {\bf 323}, 183 (2000), hep-th/9905111.
\bibitem{SooJong} 
S.~J.~Rey and J.~T.~Yee,
``Macroscopic strings as heavy quarks in large N gauge theory and  
anti-de Sitter supergravity,''
Eur.\ Phys.\ J.\ C {\bf 22}, 379 (2001)
[arXiv:hep-th/9803001];
S.~J.~Rey, S.~Theisen and J.~T.~Yee,
``Wilson-Polyakov loop at finite temperature in large N gauge theory 
and  anti-de Sitter supergravity,''
Nucl.\ Phys.\ B {\bf 527}, 171 (1998)
[arXiv:hep-th/9803135].
\bibitem{MalWilson}
J.~M.~Maldacena,
``Wilson loops in large N field theories,''
Phys.\ Rev.\ Lett.\  {\bf 80}, 4859 (1998)
[arXiv:hep-th/9803002].
\bibitem{PolRyc}
A.~M.~Polyakov and V.~S.~Rychkov,
``Loop dynamics and AdS/CFT correspondence,''
Nucl.\ Phys.\ B {\bf 594}, 272 (2001)
[arXiv:hep-th/0005173].
\bibitem{GroDru}
N.~Drukker, D.~J.~Gross and H.~Ooguri,
``Wilson loops and minimal surfaces,''
Phys.\ Rev.\ D {\bf 60}, 125006 (1999)
[arXiv:hep-th/9904191].
\bibitem{Berenstein:1998ij}
D.~Berenstein, R.~Corrado, W.~Fischler and
J.~M.~Maldacena,
``The operator product expansion for Wilson loops and
surfaces in the  large N limit,''
Phys.\ Rev.\ D {\bf 59}, 105023 (1999)
[arXiv:hep-th/9809188].
\bibitem{Semenoff}
G.~W.~Semenoff and K.~Zarembo,
``Wilson loops in SYM theory: From weak to strong
coupling,''
Nucl.\ Phys.\ Proc.\ Suppl.\  {\bf 108}, 106 (2002)
[arXiv:hep-th/0202156].
\bibitem{Wilson:1974sk}
K.~G.~Wilson,
``Confinement Of Quarks,''
Phys.\ Rev.\ D {\bf 10}, 2445 (1974).
\bibitem{Polyakov:rs}
A.~M.~Polyakov,
``Compact Gauge Fields And The Infrared Catastrophe,''
Phys.\ Lett.\ B {\bf 59}, 82 (1975).
\bibitem{KogSus}
J.~B.~Kogut and L.~Susskind,
``Hamiltonian Formulation Of Wilson's Lattice Gauge
Theories,''
Phys.\ Rev.\ D {\bf 11}, 395 (1975).
\bibitem{Giles:1977mp}
R.~Giles and C.~B.~Thorn,
``A Lattice Approach To String Theory,''
Phys.\ Rev.\ D {\bf 16}, 366 (1977).
\bibitem{BMN} D. Bernstein, J. Maldacena and H. Nastase, ``Strings in
flat space and pp waves from ${\cal N}=4$ Super Yang Mills'',
hep-th/0202021.
\bibitem{Kristjansen:2002bb}
C.~Kristjansen, J.~Plefka, G.~W.~Semenoff and M.~Staudacher,
``A new double-scaling limit of N = 4 super Yang-Mills theory and PP-wave  strings,''
Nucl.\ Phys.\ B {\bf 643}, 3 (2002)
[arXiv:hep-th/0205033].
\bibitem{G7}
N.R. Constable, D.Z. Freedman, M. Headrick, S. Minwalla, L. Motl,
A. Postnik and W. Skiba, ``PP-wave string interactions from perturbative
Yang-Mills theory'', hep-th/0205089.
\bibitem{Zhou}
J.~G.~Zhou,
``pp-wave string interactions from string bit model,''
Phys.\ Rev.\ D {\bf 67}, 026010 (2003)
[arXiv:hep-th/0208232].
\bibitem{HV}
H. Verlinde, ``Bits, Matrices and $1/N$'', hep-th/0206059; D. Vaman and
H. Verlinde, `Bit Strings from ${\cal N}=4$ Gauge Theory'', hep-th/0209215.
\bibitem{tHooft}
G.~'t Hooft,
``A Planar Diagram Theory For Strong Interactions,''
Nucl.\ Phys.\ B {\bf 72}, 461 (1974).
\bibitem{Karch:2002vn}
A.~Karch,
``Lightcone quantization of string theory duals of free field theories,''
arXiv:hep-th/0212041.
\bibitem{MT} R.R. Metsaev and A.A. Tseytlin, ``Superstring action in
AdS$_5 \times$ S$^5$: $\kappa$ -symmetry light cone gauge'', Phys. Rev. 
{\bf D63}, 046002 (2001), hep-th/0007036.
\bibitem{MTT} R.R. Metsaev, C.B. Thorn and A.A. Tseytlin, ``Light-cone
superstring in AdS space-time'', Nucl. Phys. {\bf B596}, 151 (2001),
hep-th/0009171.
\bibitem{Mack}
G.~Mack,
``All Unitary Ray Representations Of The Conformal Group SU(2,2) 
With Positive Energy,''
Commun.\ Math.\ Phys.\  {\bf 55}, 1 (1977).
\bibitem{Dobrev}
V.~K.~Dobrev and V.~B.~Petkova,
``All Positive Energy Unitary Irreducible Representations Of 
Extended Conformal Supersymmetry,''
Phys.\ Lett.\ B {\bf 162}, 127 (1985).
\bibitem{GM} M. Gunaydin and N. Marcus, ``The spectrum of the S$^5$
compactification of the chiral $N=2, D=10$ supergravity and the
unitary supermultiplets of U$(2,2|4)$'', Class. Quant. Grav. {\bf 2},
L11 (1985).
\bibitem{KRN} H.J. Kim, L.J. Romans and P. van Nieuwenhuizen, ``The
mass spectrum of chiral $N=2, D=10$ supergravity on S$^5$'',
Phys. Rev. {\bf D32}, 389 (1985).
\bibitem{SM} 
S.~Minwalla,
``Restrictions imposed by superconformal invariance
on quantum field  theories,''
Adv.\ Theor.\ Math.\ Phys.\  {\bf 2}, 781 (1998)
[arXiv:hep-th/9712074].
\bibitem{ZFF}
S.~Ferrara, C.~Fronsdal and A.~Zaffaroni,
``On N = 8 supergravity on AdS(5) and N = 4 
superconformal Yang-Mills  theory,''
Nucl.\ Phys.\ B {\bf 532}, 153 (1998)
[arXiv:hep-th/9802203].
\bibitem{Metsaev:2002vr}
R.~R.~Metsaev,
``Massless arbitrary spin fields in AdS(5),''
Phys.\ Lett.\ B {\bf 531}, 152 (2002)
[arXiv:hep-th/0201226].
\bibitem{Met} R.R. Metsaev, ``Light cone gauge formulation of IIB
supergravity in AdS$_5 \times$ S$^5$ background and AdS/CFT
correspondence'', Phys. lett. {\bf B468}, 65 (1999), hep-th/9908114.
\bibitem{Tseytlin:2002gz}
A.~A.~Tseytlin,
``On limits of superstring in AdS(5) x S**5,''
Theor.\ Math.\ Phys.\  {\bf 133}, 1376 (2002)
[Teor.\ Mat.\ Fiz.\  {\bf 133}, 69 (2002)]
[arXiv:hep-th/0201112].
\bibitem{Beisert:2002bb}
N.~Beisert, C.~Kristjansen, J.~Plefka, G.~W.~Semenoff and M.~Staudacher,
``BMN correlators and operator mixing in N = 4 super Yang-Mills theory,''
Nucl.\ Phys.\ B {\bf 650}, 125 (2003)
[arXiv:hep-th/0208178].
\bibitem{G4}
N.R. Constable, D.Z. Freedman, M. Headrick and S. Minwalla, ``Operator Mixing
and the BMN Correspondence'', hep-th/0209002.
\bibitem{Polya}
A.~M.~Polyakov,
``Gauge fields and space-time,''
Int.\ J.\ Mod.\ Phys.\ A {\bf 17S1}, 119 (2002)
[arXiv:hep-th/0110196].
\bibitem{Sundborg}
B.~Sundborg,
``The Hagedorn transition, deconfinement and N = 4
SYM theory,''
Nucl.\ Phys.\ B {\bf 573}, 349 (2000)
[arXiv:hep-th/9908001].
\bibitem{Shiraz}
S. Minwalla, ICTP lectures "Stringy
Thermodynamics in Large N Gauge Theories",
April 2003. (Based on work (to appear) with O. Aharony,
J. Marcano, S. Minwalla and M. Van Raamsdonk.)
\bibitem{KleStr}
I.~R.~Klebanov and M.~J.~Strassler,
``Supergravity and a confining gauge theory: Duality
cascades and  $\chi\,$SB-resolution of naked
singularities,''
JHEP {\bf 0008}, 052 (2000)
[arXiv:hep-th/0007191].
\bibitem{Sakita}
B.~Sakita, 
``A Quest for symmetry: Selected works of Bunji Sakita,''
{\it World Scientific Series in 20th Century Physics, Vol. 22,
1999.}; 
``Reminiscences,'' arXiv:hep-th/0006083.
\bibitem{DMWp}
A.~Dhar, G.~Mandal and S.~R.~Wadia, work in progress.
\end{thebibliography}
\end{document}